\documentclass[letterpaper,twocolumn,10pt]{article}
\usepackage{usenix}

\usepackage{url, cite}
\usepackage{xspace}

\usepackage{microtype}
\usepackage{graphicx}
\usepackage{subcaption}
\usepackage{booktabs} 
\usepackage{multirow}
\usepackage[inline]{enumitem}

\usepackage{hyperref}
\usepackage{xspace}
\usepackage{tabularx}
\usepackage{tikz}

\usepackage{url, cite}
 \usepackage{verbatim}
\usepackage{multirow}
\usepackage[makeroom]{cancel}
\usepackage{dsfont}
\usepackage{amsmath}
\usepackage{bbold}

\usepackage{tikz}  

\usepackage{tabularx}
\usepackage{color, colortbl}
\usepackage{hhline}

\usepackage{algorithm}
\usepackage[noend]{algpseudocode}
\algrenewcommand\alglinenumber[1]{\footnotesize #1}

\usepackage[small, compact]{titlesec}
\usepackage[textfont={it}, font={small,bf}]{caption}
 
\usepackage{listings}
\usepackage{xcolor}
\definecolor{commentgreen}{rgb}{0, 0.5, 0}
\def\name{{Andes}\xspace}
\newcommand{\parabf}[1]{\medskip\noindent\textbf{#1}}

\definecolor{egyptianblue}{rgb}{0.06, 0.2, 0.65}
\definecolor{grey}{rgb}{0.33, 0.33, 0.33}
\definecolor{yellow}{rgb}{1.0, 0.88, 0.21}

\def\eg{{e.g.\xspace}}

\newcommand*\circled[1]{\tikz[baseline=(char.base)]{
		\node[shape=circle,draw,inner sep=0pt] (char) {#1};}}

\usepackage{hyperref}
\hypersetup{
  colorlinks=true,      
  linkcolor=blue,       
  citecolor=magenta,    
  filecolor=cyan,       
  urlcolor=red          
}

\graphicspath{{figures/}}
\usepackage{titling}
\date{}

\newenvironment{denseenum}[1][]{
  \begin{enumerate}[topsep=2pt, partopsep=0pt, leftmargin=1.5em, #1]
    \setlength{\itemsep}{2pt}
    \setlength{\parskip}{0pt}
    \setlength{\parsep}{0pt}
}{\end{enumerate}}

\newenvironment{denseitemize}[1][]{
\begin{itemize}[topsep=2.5pt, partopsep=0pt, leftmargin=1.5em, #1]
  \setlength{\itemsep}{2.5pt}
  \setlength{\parskip}{0pt}
  \setlength{\parsep}{0pt}
}{\end{itemize}}

\begin{document}

\title{\fontsize{15}{15}{\textbf{Andes: Defining and Enhancing Quality-of-Experience \\ in LLM-Based Text Streaming Services}}}

\author{
    Jiachen Liu$^{1}$  \quad Jae-Won Chung$^{1}$  \quad Zhiyu Wu$^{1,2}$   
    \quad Fan Lai$^{2}$  \quad Myungjin Lee$^{3}$  \quad Mosharaf Chowdhury$^{1}$
    \\
    \itshape{
        $^{1}$University of Michigan \qquad
        $^{2}$UIUC \qquad
        $^{3}$Cisco Systems
    }
}






\maketitle

\begin{abstract}
Large language models (LLMs) are now at the core of conversational AI services such as real-time translation and chatbots, which provide live user interaction by incrementally \emph{streaming} text to the user.
However, existing LLM serving systems fail to provide good \emph{user experience} because their optimization metrics are not always aligned with user experience.

In this paper, we first introduce and define the notion of Quality-of-Experience (QoE) for \emph{text streaming services} by considering each user's \emph{end-to-end} interaction timeline.
Based on this, we propose \name, a QoE-aware LLM serving system that enhances user experience by ensuring that users receive the first token promptly and subsequent tokens at a smooth, digestible pace, even during surge periods.
This is enabled by \name's preemptive request scheduler that dynamically prioritizes requests at the token granularity based on each request's expected QoE gain and GPU resource usage.
Our evaluations demonstrate that, compared to state-of-the-art LLM serving systems, \name improves the average QoE by up to $4.7\times$ given the same GPU resource, or saves up to 61\% GPU resources while maintaining the same high QoE.
\end{abstract}

\thispagestyle{empty}
\pagestyle{plain} 

\section{Introduction}

\begin{figure}[t!]
   \centering 
   \begin{subfigure}[c]{0.46\textwidth}
     \includegraphics[trim=8 5 8 5,clip,width=\textwidth]{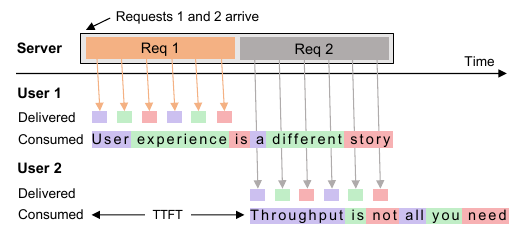}
     \caption{Existing systems significantly degraded User 2's TTFT.}\label{fig:intro-user-experience-a}
   \end{subfigure}
   \begin{subfigure}[c]{0.46\textwidth}
     \includegraphics[trim=8 5 0 0,clip,width=0.99\textwidth]{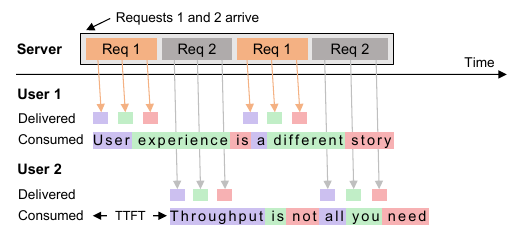}
     \caption{QoE-aware systems can schedule token generations over time. User 2's TTFT is improved without affecting User 1's token consumption.}\label{fig:intro-user-experience-b}
   \end{subfigure}
   \caption{
     Timelines for token generation, delivery, and user consumption for (a) existing and (b) QoE-aware LLM serving systems, both with capacity 1.
     Users consume tokens at their reading speed.}\label{fig:intro-user-experience}
\end{figure}

Large language models (LLMs)~\cite{gpt3-neurips20,opt-arxiv22,llama3-arxiv24,mistral-arxiv23,falcon-arxiv23,abdin2024phi,gemma2-arxiv24} have revolutionized many user-facing online applications.
Particularly, conversational AI has emerged as a dominant use case, driving over 60\% of LLM-backed applications~\cite{llm-market-report}, including chatbots, virtual assistants, language translation, and customer support. 
The meteoric rise of ChatGPT~\cite{chatgpt}, now with 300 million weekly active users~\cite{chatgpt-stats}, underscores the massive scale and demand for such services.

Conversational AI services provide \emph{interactive} conversations between the user and an autoregressive LLM~\cite{vaswani2017attention} that generates text tokens\footnote{LLMs process and generate text in units of \emph{tokens}. For instance, the word ``streaming'' may be broken down into two tokens: ``stream'' and ``ing.''} one by one.
Importantly, in order to provide a smooth conversational experience, generated tokens are incrementally streamed to the user, commonly as written text or synthesized speech.
As such, we refer to such services as \emph{text streaming services}, similar to how video streaming services stream videos to users in a frame-by-frame fashion.

User experience in video streaming is a critical and well-studied topic, where minimizing user-experienced delays (\eg, startup and buffering) during playback is crucial~\cite{video-quality-engagement-sigcomm11,video-control-plane-sigcomm12,cfa-nsdi16}.
Similarly, user experience in text streaming services hinges on
(1) the timely delivery of initial tokens, and
(2) the delivery of subsequent tokens in a smooth and digestible pace (\S\ref{sec:motivation-tss-ux}).

Existing LLM serving systems~\cite{orca-osdi22,vllm-sosp23,distserve-osdi24,sarathi-serve-osdi24,loongserve-sosp24}, however, fail to provide good user experience (\S\ref{sec:motivation-misalignment}).
During periods of request load surge, simplistic first-come-first-served (FCFS) scheduling policies adopted by existing systems cause head-of-line blocking.
This inflates time-to-first-token (TTFT) and initial user wait times, as shown in Figure~\ref{fig:intro-user-experience-a}.


We observe that existing systems generate and deliver tokens to users at a pace faster than their typical reading/listening speed.
This does not improve user experience because the token consumption speed of users is capped at their reading or listening speed.
This opens up the opportunity to control the generation and delivery of each token in each request to ensure a smooth and digestible pace, which not only enhances user experience but also creates opportunities to serve more users simultaneously with the same amount of hardware resource (\S\ref{sec:motivation-opportunities}).
As shown in Figure~\ref{fig:intro-user-experience-b}, we can redistribute computation resources across requests over time at token granularity to improve TTFT without affecting the token consumption timeline of other users.

Existing LLM serving systems have so far failed to recognize this opportunity because there is a fundamental \emph{misalignment} between user experience and optimization metrics used by existing systems. 
Server-centric metrics (\eg, token generation throughput~\cite{orca-osdi22,vllm-sosp23}) or simple statistics derived from a subset of token delivery timestamps (\eg, average/P90/P99 time-per-output-token or TTFT~\cite{distserve-osdi24,sarathi-serve-osdi24,loongserve-sosp24}) fail to fully capture user experience during the whole interactive session.
To accurately reflect user experience, every token matters!

To bridge this gap, in this paper, we first formally define the notion of Quality-of-Experience (QoE) as an optimization metric for LLM-based text streaming services (\S\ref{sec:overview-qoe}).
This is challenging in itself, as it needs to capture the user's end-to-end interaction timeline and reflect our intuitions of good and bad user experiences.
For instance, long pauses before and during text streaming should degrade the right amount of QoE, while generating tokens at a rate faster than the user's consumption speed should not improve QoE.

Based on our definition of QoE, we design \name, a QoE-aware LLM serving system.
\name co-designs the LLM inference server and the text streaming service client (\S\ref{sec:overview-architecture}).
On the server side, \name adopts a preemptive request scheduler that operates at the token granularity to optimize QoE (\S\ref{sec:scheduling}).
Designing such a scheduler is challenging:
\begin{denseenum}[label=(\alph*)]
  \item \textbf{Diverse and unpredictable resource demand.}
  Requests arrive dynamically with varying prompt lengths, response lengths, and QoE parameters (\eg, user reading speed). 
  This dynamism and diversity preclude the adoption of simple one-size-fits-all scheduling policies.
  \item \textbf{Interdependent aspects in user experience.}
  On the one hand, we want to minimize the initial waiting time for users by serving more requests in parallel that maximizes GPU memory utilization. 
  On the other hand, serving with a larger batch size may slow down token generation, potentially failing to meet user's ideal token delivery speed.
  With request input lengths and QoE parameters varying widely over time, this is critical but challenging to control.
  \item \textbf{Token-level preemption overhead.}
  While token-level request preemption is promising for QoE optimization (Figure~\ref{fig:intro-user-experience}), such fine-grained scheduling introduces additional overhead that may degrade system throughput. This may in turn degrade the QoE of every request.
\end{denseenum}

To handle these challenges, \name continuously monitors the attained QoE and resource usage of each request and dynamically prioritizes requests that are at risk of degrading their QoE.
By accounting for both memory and compute constraints, \name ensures a comprehensive optimization of the overall QoE.
Moreover, \name modulates the frequency of decision-making and incorporates the overhead of request preemption and restart in its scheduling decision.

\name's server implements push-based streaming, which immediately transmits tokens to the client as they are generated.
This is because the server is typically resource-constrained, and generated tokens are delivered exclusively to the user who submitted the request.
To provide a smooth streaming experience at a consistent pace, on the client side, \name introduces a \emph{token pacer} that temporarily buffers excess tokens generated by the server and delivers them to the user precisely at the user's consumption speed (\S\ref{sec:implementation}).

We evaluate \name with popular LLMs with various sizes (3.8B to 70B) and architectural characteristics (Dense and Mixture-of-Experts, Multi-Head Attention and Grouped-Query Attention) on three different datasets with varying input and output sequence lengths (\S\ref{sec:eval}).
Compared with state-of-the-art LLM serving systems (vLLM~\cite{vllm-sosp23} and Sarathi-Serve~\cite{sarathi-serve-osdi24}), \name improves average QoE by up to $4.7\times$ given the same GPU resource, or provides up to 61\% GPU savings while maintaining the same high QoE.

Overall, we make the following contributions:
\begin{denseitemize} 
  \item We identify a misalignment between the user experience of LLM-based text streaming services and optimization metrics pursued by state-of-the-art LLM serving systems.
  \item We formally define QoE for text streaming services that fully captures their user experience.
  \item We design and implement \name, an LLM serving system that co-designs the server (token-level preemptive request scheduler) and the client (token pacer).
  \item We evaluate \name on diverse workloads and show that it significantly improves QoE and saves GPU resources.
\end{denseitemize}

\begin{figure}[t]
  \centering
  \begin{subfigure}[c]{0.23\textwidth}
    \includegraphics[trim=21 19 0 0,clip,scale=0.30]{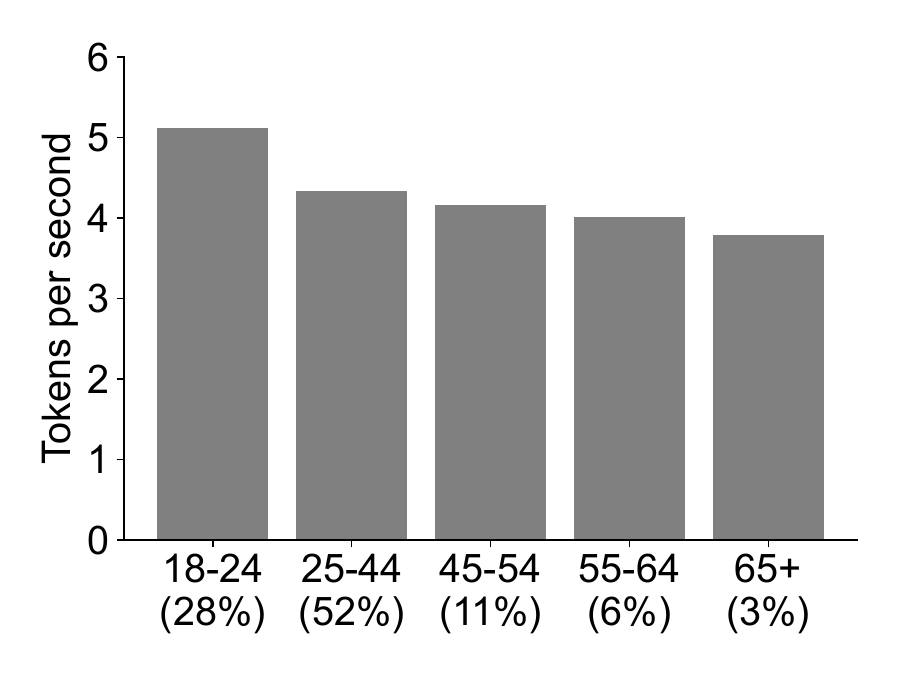} 
    \caption{Reading speed and age group}\label{fig:expected-tds-age}
  \end{subfigure}
  \hfil
  \begin{subfigure}[c]{0.23\textwidth}
    \includegraphics[trim=39 19 0 0,clip,scale=0.30]{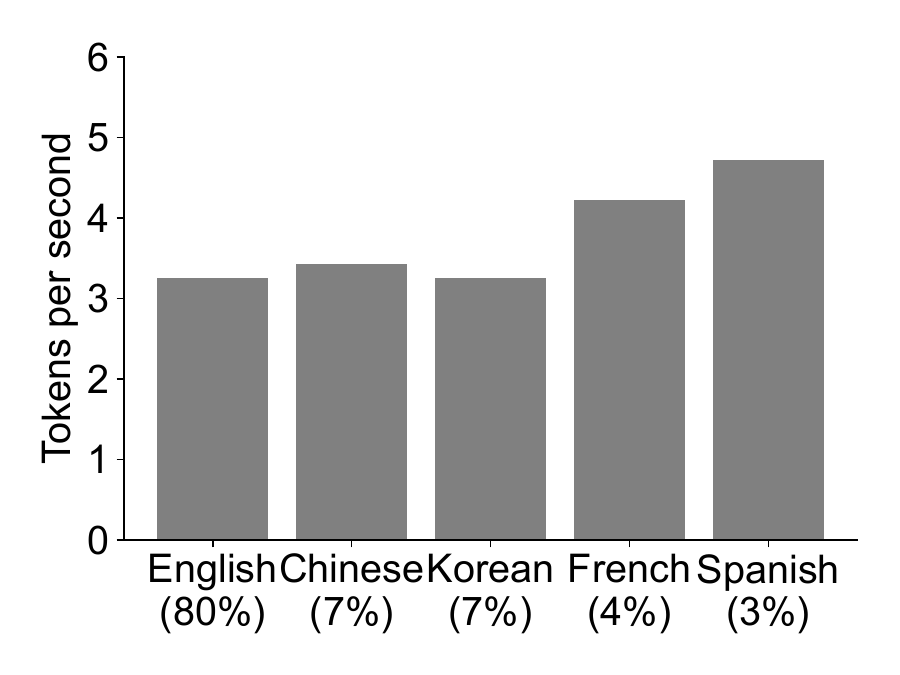} 
    \caption{Speaking speed and language} 
  \end{subfigure}
  \caption{
    User token consumption speed based on demographic group, counting one word as 1.3 tokens on average~\cite{openai-tokens}.}\label{fig:expected-tds}
\end{figure}

\section{Background and Motivation}\label{sec:motivation}

In this section, we first introduce LLM-based text streaming services and their user experience (\S\ref{sec:motivation-tss-ux}).
We then discuss the limitations of existing LLM serving systems (\S\ref{sec:motivation-misalignment}) and opportunities for improving user experience (\S\ref{sec:motivation-opportunities}).

\subsection{Text Streaming Services and User Experience}\label{sec:motivation-tss-ux}

Conversational AI services including chatbots, virtual assistants, and real-time translation aim to provide a smooth conversational experience by incrementally streaming text to users, either in visual text or audible synthesized speech, instead of having them wait for tens of seconds before the whole response is generated.

The user's interaction timeline with such text streaming services can be divided into
(1) the \emph{initial waiting phase} and
(2) the \emph{token consumption phase}.
During the former phase, time-to-first-token (TTFT) matters; the first token should be delivered to users before they lose patience.
The target TTFT should depend on the service, be it constant or proportional to the prompt length.
During the latter phase, users expect the server's token delivery speed (TDS) to match their consumption speed (Figure~\ref{fig:expected-tds}).
More specifically, a slower TDS hurts user experience as users will perceive the service as slow or lagging, but a TDS faster than the user's consumption speed does not improve user experience.

\subsection{Existing Systems Provide Poor User Experience}\label{sec:motivation-misalignment}

\begin{figure}[t]
  \centering 
  \includegraphics[trim=10 20 10 0,clip,width=0.47\textwidth]{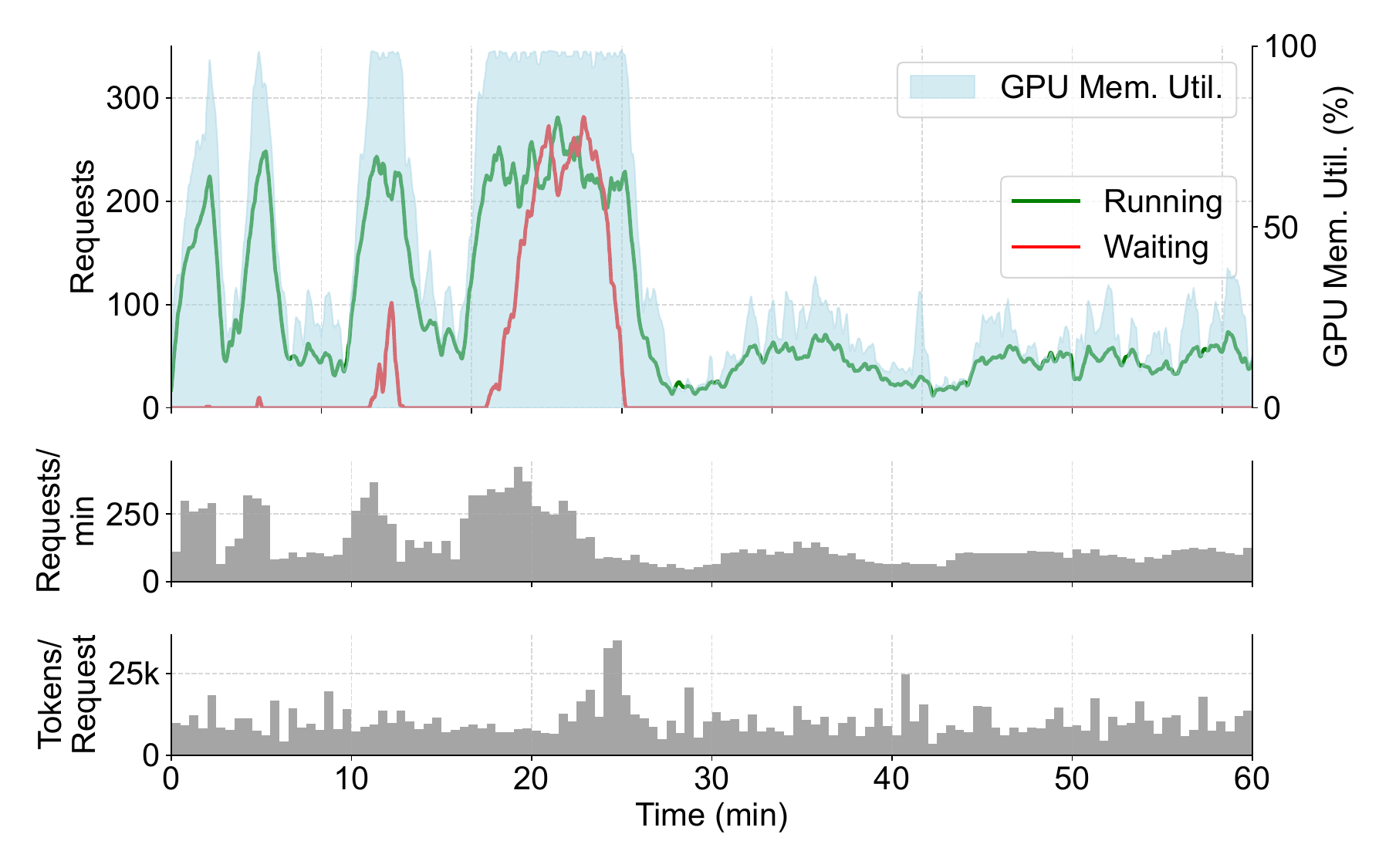}     
  \caption{
    vLLM serving requests from BurstGPT.\@
    During load surges, vLLM builds up a very long queue due to head-of-line blocking.
    Still, under normal load, GPU memory is underutilized.}\label{fig:burst-fcfs}
\end{figure}

Existing LLM serving systems~\cite{orca-osdi22,vllm-sosp23,distserve-osdi24,sarathi-serve-osdi24} fail to deliver good user experience for text streaming services.
To gain deeper insight into how existing LLM serving systems behave with real-world text streaming services, we replay a one-hour slice of BurstGPT~\cite{burstgpt-arxiv24}, a real-world LLM serving request trace, with vLLM~\cite{vllm-sosp23} serving Phi-3.5-MoE 16$\times$3.8B~\cite{abdin2024phi} on 8 $\times$ A100 GPUs.
Figure~\ref{fig:burst-fcfs} (first row) shows vLLM's GPU memory utilization and the number of running and waiting requests over time.
Above all, we can observe large and frequent load surges caused by spikes in both request rate (second row) and the number of tokens in requests (third row).
During surge periods, vLLM's first-come-first-served (FCFS) scheduling policy -- adopted by most existing LLM serving systems as well -- causes significant head-of-line blocking and queuing delay.
This leads to an average TTFT of 10.4 seconds, which is likely beyond the patience limit of most users~\cite{web-loading-speed-google18}.
On the other hand, vLLM's average TDS is 11.2 tokens/s, a rate that significantly exceeds any reasonable user token consumption speed (Figure~\ref{fig:expected-tds}).
This does not enhance user experience, as the timeline of users consuming tokens does not change even when tokens are delivered faster than their consumption speed.

To understand this further under varying degrees of load surges, we create synthetic 20-minute request traces that contain a load surge period with varying durations of bursts (See Section~\ref{sec:e2e-cyclic} for more details).
Figures~\ref{fig:burst-ttft} and~\ref{fig:burst-tds} show the average TTFT and average TDS of vLLM serving Phi-3.5-MoE 16$\times$3.8B, respectively.
Consistent with the real BurstGPT trace, we observe:
(1) TTFT is significantly inflated due to head-of-line blocking, even under moderate load surges, and
(2) TDS consistently exceeds any reasonable user token consumption speed, even under severe load surges.

\subsection{Opportunities for Improving User Experience}\label{sec:motivation-opportunities}

\begin{figure}[t]
  \centering
  \begin{subfigure}[b]{0.23\textwidth}
    \includegraphics[trim=0 0 0 0,clip,width=\textwidth]{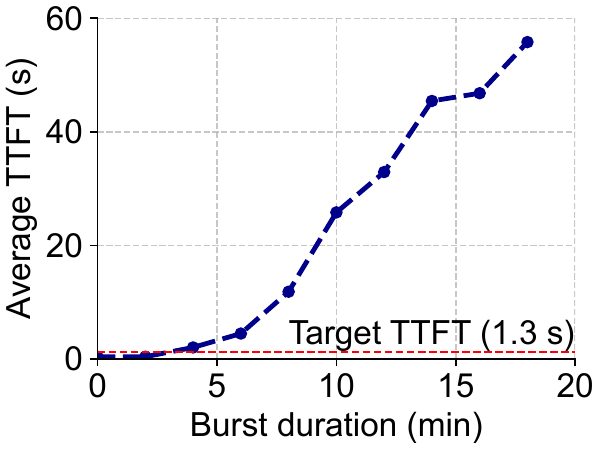}
    \caption{Average TTFT}\label{fig:burst-ttft}
  \end{subfigure}
  \begin{subfigure}[b]{0.23\textwidth} 
    \includegraphics[trim=0 0 0 0,clip,width=\textwidth]{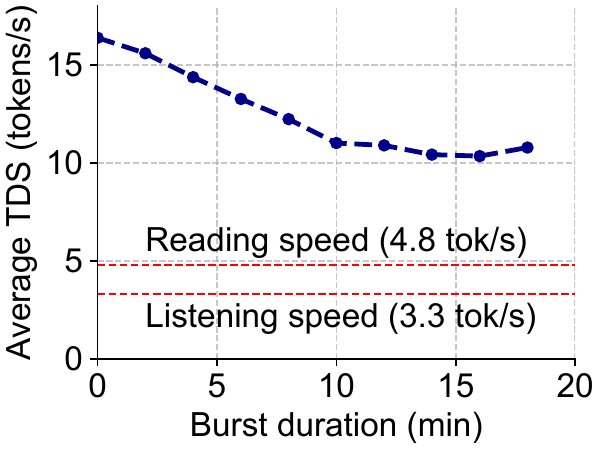}
    \caption{Average TDS}\label{fig:burst-tds}
  \end{subfigure}
  \caption{
    vLLM's average TTFT and TDS while varying the duration of the load surge in our 20-minute trace.
    TTFT target is set to 1.3 s, as recommended by Google for web page loading~\cite{web-loading-speed-google18}.
    User reading/listening speeds are 4.8 and 3.3 tokens/s, derived from Figure~\ref{fig:expected-tds}.
    TTFT is inflated significantly, whereas TDS remains excessively fast compared to any token consumption speed.}
    \label{fig:burst-ttft-tds}
\end{figure}


In essence, Section~\ref{sec:motivation-misalignment} reveals a \emph{misallocation} of computational resources over time -- instead of continuing to allocate compute to requests that have generated enough tokens for their users to consume, it makes more sense to redistribute compute to requests that have not generated enough (or, \emph{any}) tokens for their users.
This insight motivates a preemptive scheduling approach that can reduce TTFT inflation caused by head-of-line blocking, allowing more requests to be served concurrently. 
Preemption and resumption overhead may reduce the average token generation speed of individual requests, but as long as it remains faster than the user's token consumption speed, user experience will not be affected.

How much is the potential gain under ideal circumstances?
Opportunity fundamentally comes from the gap between the server's excessive token generation speed and the user's token consumption speed. 
Instead of serving the same set of requests to completion and over-generating tokens for them, an ideal preemptive scheduler could switch to different requests back and forth while ensuring that each request generates tokens at a rate that matches the user's consumption speed.
For instance, for the workload in Figure~\ref{fig:burst-ttft-tds}, the server generates 11 tokens/s under moderate load surges (with burst duration 10 minutes), while the user can only consume 4.8 tokens/s.
Therefore, each request just needs to be served for 1 second every $11 / 4.8 = 2.3$ seconds, alternatively allowing the server to serve $2.3\times$ more requests concurrently.
This is an ideal upper limit estimation assuming no scheduling, prefill, preemption, and restoration overhead, but we show in Section~\ref{sec:eval} that \name can still realize a significant portion of the ideal gains.

\section{\name Overview}\label{sec:overview}

\name is an LLM serving system for text streaming services that enhances user experience by co-designing the server and the client.
In this section, we introduce a formal definition of Quality-of-Experience (QoE) for text streaming services (\S\ref{sec:overview-qoe}), and provide an overview of \name's architecture and request lifecycle designed to optimize QoE (\S\ref{sec:overview-architecture}).

\subsection{Quality-of-Experience in Text Streaming}\label{sec:overview-qoe}

\begin{figure}[t]
  \centering 
  \begin{subfigure}[c]{0.49\textwidth}
     \includegraphics[trim=0 0 0 10,clip,width=0.95\linewidth
     ]{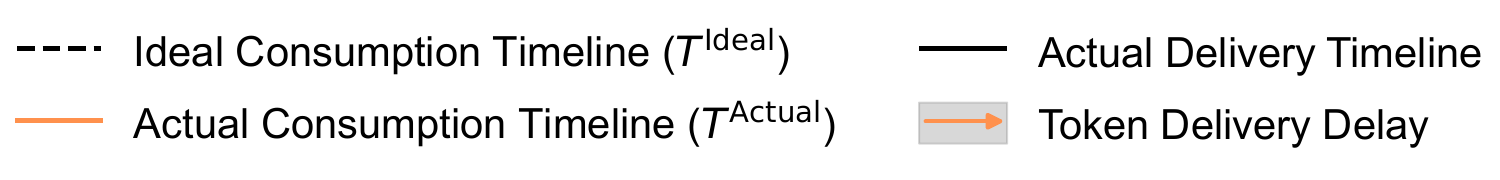} 
  \end{subfigure} 

  \begin{subfigure}{0.23\textwidth}
    \centering
    \includegraphics[trim=15 15 15 15,clip,scale=0.39]{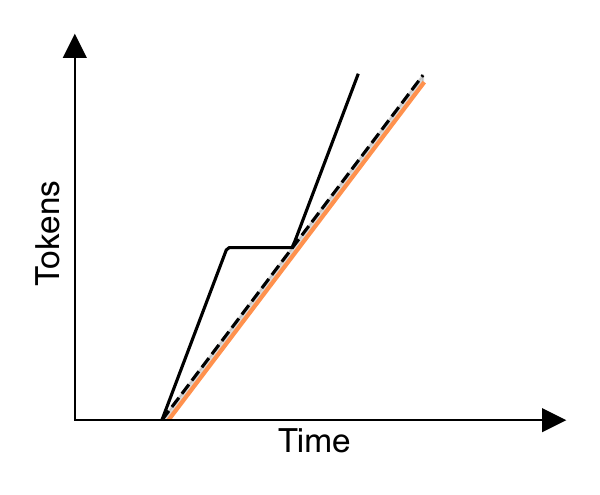} 
    \caption{Perfect experience.}\label{fig:perfect-qoe}
  \end{subfigure}%
  \hfil%
  \begin{subfigure}{0.23\textwidth}
    \centering
    \includegraphics[trim=15 15 15 15,clip,scale=0.39]{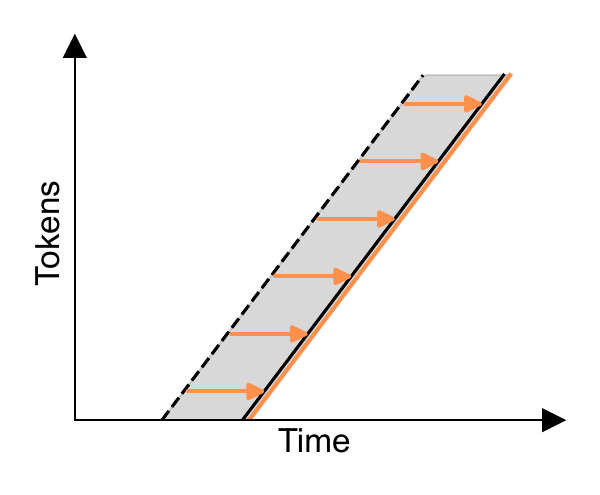} 
    \caption{Delay in first token.}\label{fig:ttft-missed}
  \end{subfigure}%

  \begin{subfigure}{0.23\textwidth}
    \centering
    \includegraphics[trim=15 15 15 15,clip,scale=0.39]{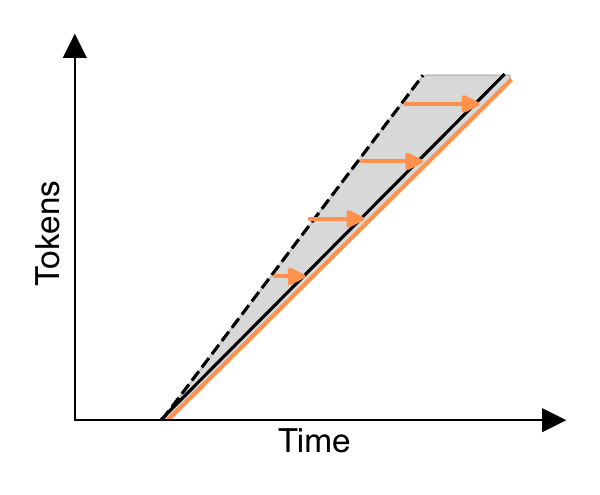} 
    \caption{Delay in every subsequent token.}\label{fig:tds-missed}
  \end{subfigure} 
  \hfil%
  \begin{subfigure}{0.23\textwidth}
    \centering
    \includegraphics[trim=15 15 15 15,clip,scale=0.39]{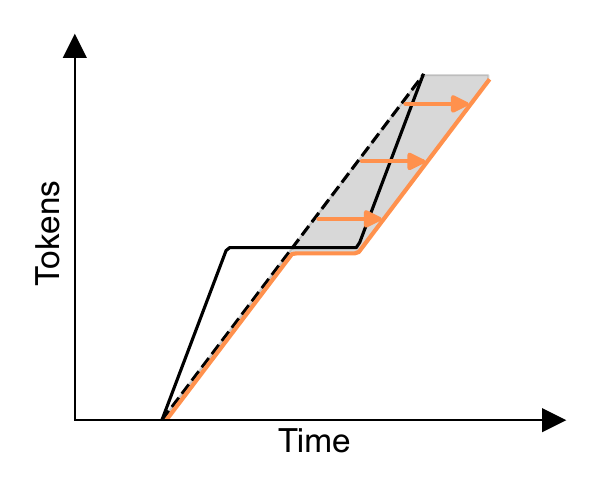} 
    \caption{Long pause in the middle.}\label{fig:pause-in-middle}
  \end{subfigure}  

  \caption{User experience examples.
    Users expect tokens to be delivered along the fixed Ideal Consumption Timeline, with its slope being their reading/listening speed.
    (a) Without any Token Delivery Delays, users' Actual Consumption Timeline aligns with their Ideal Consumption Timeline.
    (b, c, d) However, when tokens are delayed, users perceive the delay and user experience degrades.}\label{fig:qoe}
\end{figure}

Existing LLM systems have thus far overlooked the opportunity to enhance user experience (\S\ref{sec:motivation}).
Fundamentally, this is because widely used system optimization metrics only consider a limited portion of the user's \emph{token consumption timeline}, failing to fully capture user experience.
Server-centric metrics like token generation throughput ignore the token delivery timeline of individual requests.
TTFT only captures the user's consumption of the first token, and average/P90/P99 time-per-output-token (TPOT) can miss outlier TPOT inflations that lead to user-perceived pauses in text streaming.

As such, we are in need of a QoE metric that fully captures user experience in text streaming.
Figure~\ref{fig:qoe} shows four foundational cases that guide the design of QoE.
The Ideal Consumption Timeline represents the user's ideal experience, with low TTFT and every subsequent token being delivered precisely at the user's consumption speed (\eg, reading speed).
\begin{denseenum}
  \item Figure~\ref{fig:perfect-qoe} (\emph{Perfect user experience}): 
  \looseness=-1
  The server delivered every token no later than the user's expectation, allowing users to consume tokens following the ideal timeline. 
  Delivering tokens earlier than the ideal timeline does not improve user experience, as users cannot consume text faster.

  \item Figure~\ref{fig:ttft-missed} (\emph{Long initial delay}): 
  The request stayed in the server's queue for a long time due to head-of-line blocking.
  As a result, the user experienced a long initial wait time, followed by subsequent tokens delivered at the user's reading speed. 
  Delays in earlier tokens result in a cascading delay in every subsequent token.

  \item Figure~\ref{fig:tds-missed} (\emph{Slow streaming speed}):
  When the server processes a larger batch of requests, its token generation latency increases, potentially failing to meet the user's ideal token consumption speed.
  In this case, the first token was delivered on time, but the user experienced delays in every subsequent token.

  \item Figure~\ref{fig:pause-in-middle} (\emph{Pause during streaming}): 
  The server preempted the request to avoid OOM errors, pausing its token delivery in the middle.
  When the Actual Delivery Timeline crosses the Ideal Consumption Timeline, the user runs out of tokens to read and experiences a pause. 
  This case is particularly insidious, as TTFT or average TPOT does not degrade; both the first and last tokens were delivered on time, hiding the long pause in the middle.
\end{denseenum}

As can be seen above, we can gauge user experience (and the degradation thereof) based on how much the user's Actual Consumption Timeline ($T^\textrm{Actual}$) deviated from the Ideal Consumption Timeline ($T^\textrm{Ideal}$).
Thus, we define:
\begin{equation}
  S_\textrm{delay} = \sum_{i=1}^{n} {\left( T_i^\textrm{Actual} - T_i^\textrm{Ideal} \right)},
\end{equation}
where $T_i^\textrm{Actual}$ and $T_i^\textrm{Ideal}$ respectively denote token $i$'s actual and ideal consumption timestamp, and $n$ is the total number of tokens consumed by the user.
$S_\textrm{delay}$ measures how much the Actual Consumption Timeline deviated from the Ideal Consumption Timeline (gray shaded area in Figure~\ref{fig:qoe}), and properly reflects the delay of earlier tokens creating cascading delays in later tokens.
However, as more tokens are generated for the request, user experience degradation from earlier delays ought to be diluted.
Thus, we introduce a normalizer:
\begin{equation} 
  S_\textrm{whole} = \sum_{i=1}^{n} {\left( T_n^\textrm{Actual} - T_i^\textrm{Ideal} \right)}.
\end{equation}
$S_\textrm{whole}$ covers $S_\textrm{delay}$ and the area below the Actual Consumption Timeline, always being larger than $S_\textrm{delay}$.
With this, our QoE definition is:
\begin{equation}
  \textrm{QoE} = 1 - \frac{S_\textrm{delay}}{S_\textrm{whole}}.
  \label{eq:qoe}
\end{equation}
With no token delay at all, $S_\textrm{delay}$ will be zero, leading to a perfect QoE of 1.
On the other hand, when tokens are delayed, $S_\textrm{delay}$ will take up a larger proportion inside $S_\textrm{whole}$, reducing the value of QoE.
Finally, when no tokens arrive, $S_\textrm{delay}$ will equal $S_\textrm{whole}$, leading to the worst possible QoE of 0.

QoE can be computed on requests in any state, be it queued, running, or finished.
The optimization objective of \name is to maximize QoE across all requests.

\subsection{\name Architecture}\label{sec:overview-architecture}

\begin{figure}[t]
  \centering
  \includegraphics[trim=0 0 0 0,clip,width=0.98\columnwidth]{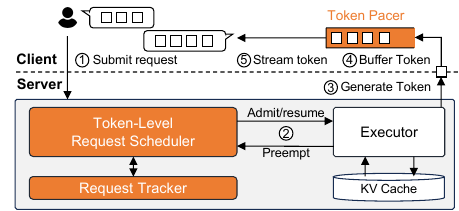} 
  \caption{High-level architecture of \name. Components designed by \name are colored in orange.}\label{fig:overview}
\end{figure}

\paragraph{System Components.}
\name co-designs the LLM inference server and the application client.
Figure~\ref{fig:overview} shows the high-level architecture of \name.
On the server side, the \textbf{Request Tracker} maintains the control state of each request, including its QoE parameters (TTFT target and user token consumption speed), prompt and partial response, timestamps of each generated token, and resource usage.
Based on this information, the \textbf{Token-Level Request Scheduler} (\S\ref{sec:scheduling}) makes runtime decisions on which requests to admit/resume or preempt.
Such decisions are carried out by the data plane (\textbf{Executor} and \textbf{KV Cache}), which is also responsible for running LLM inference and generating tokens.
The server implements push-based streaming, and the client-side \textbf{Token Pacer} (\S\ref{sec:implementation}) smooths out the pace of incoming tokens by temporarily buffering tokens and delivering them to the user at their consumption speed.
The server is fully aware of the Token Pacer, and will generate and push just enough tokens so that the Token Pacer does not run out of tokens to deliver.

\paragraph{Request Lifecycle.}
First, a request is \circled{1} submitted by the user through the application-integrated client, and the client also informs the Token Pacer of the user's QoE parameters.
The request is then enqueued into the server, and the Request Tracker initializes and continuously tracks the request's status.
During its lifetime, the request can either be waiting (enqueued or preempted) or running.
\circled{2} When the Token-Level Request Scheduler decides to preempt a running request, it will transition into the waiting state after saving its intermediate state.
On the other hand, a waiting request begins running when the scheduler decides to either newly admit or resume it, restoring its intermediate state if any.
When a request is running, \circled{3} the Executor generates tokens and \circled{4} pushes them immediately to its corresponding user's Token Pacer.
Regardless of the request's state in the server, the Token Pacer \circled{5} drains buffered tokens and delivers tokens to users following their Ideal Consumption Timeline.

\section{QoE-Aware Token-Level Scheduler}\label{sec:scheduling}

We dive into how \name optimizes the QoE of text streaming services by performing token-level preemptive request scheduling.
We start by formulating the scheduling problem assuming that there is no preemption overhead (\S\ref{sec:scheduling-problem-formulation}), as the overhead itself depends on the scheduling decisions dictated by the scheduling policy. 
We then propose an efficient scheduling algorithm for the problem (\S\ref{sec:scheduling-solution-design}) and refine the solution to incorporate preemption overhead (\S\ref{sec:scheduling-tradeoff}).

\subsection{Problem Formulation}\label{sec:scheduling-problem-formulation}

We first discuss the objective and constraints of \name, and then put them together to formulate the scheduling problem.

\paragraph{Scheduling Setup and Objective.}
\name's scheduler operates in an online setting where user requests arrive over time with diverse input lengths and QoE parameters. 
Its objective is to maximize the \emph{average QoE} across all requests.\footnote{Alternative objectives can also be used. See Appendix~\ref{sec:apdx-other-objectives}.}  

Like any other online serving system, it is very difficult, if not impossible, to perfectly plan execution into the future because the arrival time, input and output lengths, and QoE parameters of each request are not known in advance.
Instead, among ongoing (waiting and running) requests, \name decides which requests to serve at the beginning of each scheduling quantum.
Based on this decision, requests are admitted/resumed and preempted as needed.

\name decides whether or not to serve a request based on the \emph{QoE gain} it is expected to bring when it is served compared to when it is not served, calculated as:
\begin{equation}
  Q_{\textrm{serve}, i} - Q_{\textrm{wait}, i}
  \label{eq:scheduling-objective-value}
\end{equation}
where $Q_{\textrm{serve}, i}$ and $Q_{\textrm{wait}, i}$ are request $i$'s QoE when it is served and not served, respectively.
As we are uncertain about whether a request will be continued to be served or preempted in the future, we estimate the QoE gain of a request in the upcoming time frame of length $\Delta t$.
We evaluate the impact of $\Delta t$ on \name in Section~\ref{sec:ablation}.
Naturally, in order to maximize the average QoE, we would want to serve more of the requests that will bring large QoE gains.

\paragraph{Resource Constraints.}
LLM serving systems are bottlenecked primarily by two GPU resources: memory and compute.
These impose constraints on which requests can be concurrently served by the system.

First, each token in a request's context (both input and output tokens) consumes one entry in the LLM serving system's KV cache~\cite{gpt3-neurips20}.
As GPU memory is limited, there exists an upper limit on the number of KV cache entries the GPU can hold together with model weights and intermediate tensors.
The total KV cache size of requests that are served must not cross this upper limit.

In addition to memory constraints, \name must also consider compute constraints, which affects the computation latency of token generation by the executor; a larger batch size generally increases computation latency.\footnote{More precisely, token generation latency is a function of batch size and the total number of tokens in the batch, but batch size and total number of tokens are nearly perfectly correlated, allowing us to eliminate the latter and only leave batch size. See Appendix~\ref{sec:apdx-modeling-token-generation-latency} for a more detailed analysis.}
Thus, while a large batch size $B$ serves more requests concurrently, it will also increase the latency to generate one token from the perspective of each request.
This may lead to individual requests generating tokens too slowly and degrading QoE.\@
On the other hand, a smaller batch size would lead to faster token generation, but the server is serving fewer requests, potentially degrading the QoE of those that are left waiting.


\begin{figure}[t]
    \centering
    \begin{subfigure}[b]{0.47\textwidth}
      \centering
      \includegraphics[trim=0 0 0 0,clip,width=0.95\linewidth]{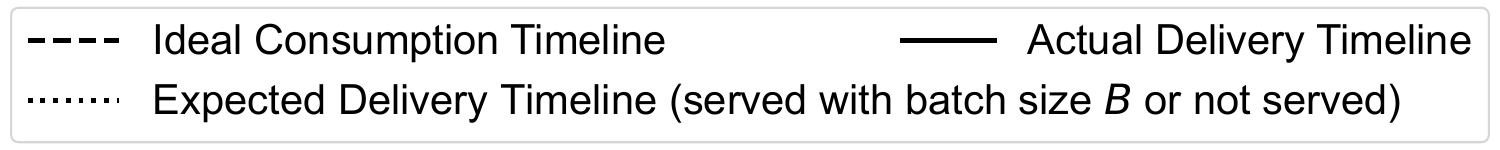} 
    \end{subfigure}

    \begin{subfigure}[b]{0.23\textwidth}
      \centering
      \includegraphics[trim=20 20 20 0,clip,width=0.99\columnwidth]{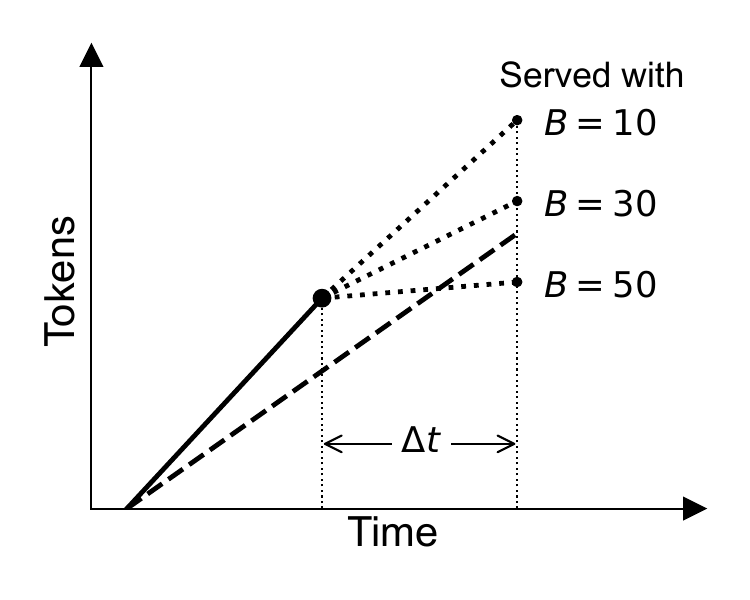} 
      \caption{$Q_{\textrm{serve}, i}(B)$}\label{fig:scheduling-qoe-q-serve}
    \end{subfigure}%
    \hfill
    \begin{subfigure}[b]{0.23\textwidth}
      \centering
      \includegraphics[trim=20 20 20 0,clip,width=0.99\columnwidth]{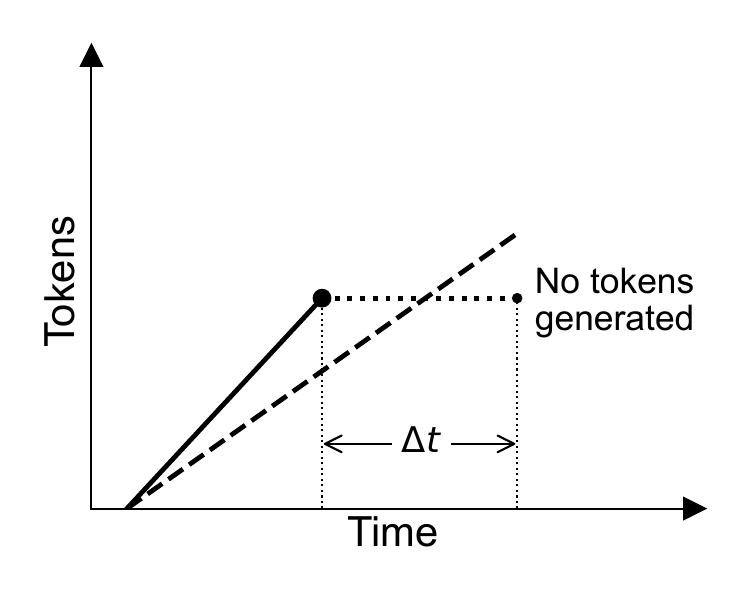} 
      \caption{$Q_{\textrm{wait}, i}$}\label{fig:scheduling-qoe-q-wait}
    \end{subfigure}
    \caption{Visualization of $Q_{\textrm{serve}, i}(B)$ and $Q_{\textrm{wait}, i}$. The former depends on batch size $B$ whereas the latter is a constant. With batch size 50, request $i$ no longer has perfect QoE.}\label{fig:scheduling-qoe}
\end{figure}

As such, determining the right batch size $B$ is critical in maintaining the right token generation speed for requests, which in turn affects QoE gain of serving each request.
However, \name would know the batch size of the upcoming scheduling quantum only after making its scheduling decision.
Fortunately, \name can still \emph{estimate} the QoE gain of each request for any batch size $B$, and use this to make the scheduling decision.
Figure~\ref{fig:scheduling-qoe} provides an example of this.
When batch size is small ($B=10$), tokens are generated quickly and the request maintains perfect QoE.\@
However, as $B$ increases ($B=30$ and $B=50$), token generation slows down due to higher computation load, and perfect QoE will be broken in the case of $B=50$.
On the other hand, when the request is not served and waiting, it does not generate any tokens and therefore batch size does not affect its QoE.\@


\paragraph{Problem Formulation.}
Putting these together, we have the following scheduling problem:
\begin{equation}
	\begin{aligned}
    \max_{x} \quad & \sum_{i=1}^{N} \left( Q_{\text{serve}, i}(B) - Q_{\text{wait},i} \right) \cdot x_i \\
    \textrm{s.t.} \quad & x_i \in \{ 0, 1 \}, ~ i \in 1, \ldots, N \\
                        & \sum_{i=1}^{N} x_i = B \\
                        & \sum_{i=1}^N l_i x_i \le M \\
	\end{aligned}
  \label{eq:scheduling-optimization-problem}
\end{equation} 
where $N$ is the total number of ongoing requests, and for request $i$, $l_i$ is its context length, and $x_i$ equals 1 if the request will be served and 0 if not.
The objective has been updated so that QoE gain properly depends on batch size $B$.
The second constraint enforces that batch size should be exactly $B$, and the third ensures that the total context length in the batch does not exceed the GPU's memory.
Notice that batch size $B$ is treated as a given; the optimization problem in Equation~\ref{eq:scheduling-optimization-problem} has to be solved for $\forall B \in [1, N]$ and the $x$ that leads to the largest optimum across $B$ will be chosen.

Given the problem formulation, we observe that it resembles that of the classic knapsack problem~\cite{kellerer2004knapsack}.
The goal is to select items (requests) to put in a knapsack (inference server) so that total item value (QoE gain) is maximized and total weight (context length) does not exceed the knapsack's capacity (memory capacity).
Yet, the fact that the value of each item (QoE gain) depends on how many items end up in the knapsack (batch size) makes it a harder variant.

\paragraph{Problem Hardness.}
The problem in Equation~\ref{eq:scheduling-optimization-problem} is weakly NP-Hard~\cite{kellerer2004knapsack}. 
3D dynamic programming (DP) can solve the problem optimally in pseudo-polynomial time $O(M N^2)$ (See Appendix~\ref{sec:apdx-3d-dp}), which is likely too slow as the number of requests $N$ and the maximum number of tokens that can fit in memory $M$ are easily in the order of hundreds and thousands, respectively.
Furthermore, Equation~\ref{eq:scheduling-optimization-problem} has to be solved for $\forall B \in [1, N]$, which is clearly intractable.

\subsection{Priority-Based QoE-Aware Scheduling}\label{sec:scheduling-solution-design}

It is computationally prohibitive to solve the problem introduced in the previous section as is.
Therefore, in this section, we propose an approximate but efficient solution.

\paragraph{A Motivating Example.}
We propose a \emph{priority}-based greedy heuristic that approximates the solution to the knapsack problem. 
The algorithm assigns a priority to each request based on its potential QoE gain relative to its resource usage.
Intuitively, requests that require less GPU resource but still bring large QoE gain should be favored, allowing the system maximize average QoE under resource constraints.

\begin{figure}[t]
  \centering
  \includegraphics[trim=0 79 100 35,clip,width=0.95\columnwidth]{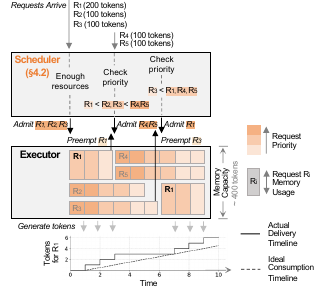} 
  \caption{Toy example of \name's token-level preemptive request scheduling.
    The priority of each request changes over time, as indicated by colors.
    Scheduling decisions are made based on this dynamic priority and the amount of available GPU resources.}\label{fig:solver-overview}
\end{figure}

We illustrate the scheduling process with a toy example to demonstrate the system's behavior as in Figure~\ref{fig:solver-overview}, along with the token delivery process of one request $R_1$ (bottom row).
At $t=0$, three requests $R_1$, $R_2$, and $R_3$ with different GPU memory demands -- represented by the number of tokens in the prompt -- arrive and start running.
Even at $t=0$, request $R_1$ is assigned a lower priority than $R_2$ and $R_3$ because it has a longer context length and thus consumes more GPU resources.
All three requests generate one token per iteration, as also reflected by request $R_1$'s token timeline.
At $t=3$, two new requests $R_4$ and $R_5$ arrive.
However, the executor is fully occupied, so the scheduler needs to decide which requests to preempt and which requests to admit.
At the moment, $R_1$ has the lowest priority because (1) it has accumulated enough tokens for the user to consume, leading to a lower potential QoE gain, and (2) it consumes the most amount of GPU memory.
Therefore, the scheduler preempts $R_1$ temporarily and admits $R_4$ and $R_5$. 
Even after $R_1$ gets preempted, its Actual Delivery Timeline curve is still above the Ideal Consumption Timeline, meaning the user will not experience any pause.
At $t=6$, the Actual Delivery Timeline of $R_1$ is about to drop below the Ideal Consumption Timeline, so the scheduler preempts $R_3$ with the lowest priority and resumes $R_1$.
This allows $R_1$ to resume token generation, ensuring the user continues to receive tokens without experiencing any pause.

\paragraph{Priority-Based Greedy Packing.}
Based on this intuition, we define request $i$'s priority as:
\begin{equation}
  \frac{Q_{\textrm{serve}, i}(B) - Q_{\textrm{wait}, i}}{l_i}.
  \label{eq:scheduling-priority}
\end{equation}
This priority function meets our design goals:
\begin{denseitemize}
  \item Requests with higher QoE gain and lower GPU resource usage will be prioritized.

  \item A request with long context length ($l_i$) will be preempted first, freeing enough GPU memory to potentially bring in \emph{more than one} waiting requests.\footnote{The overhead of preemption depends on how much memory was freed, not the number of requests. Therefore, for the same amount of memory freed from preemption, it's better to free a smaller number of requests.}
  This reduces the number of preemptions required to alleviate head-of-line blocking. 

  \item As a request receives service, its context length ($l_i$) will increase, automatically deprioritizing itself.
  On the other hand, the QoE gain of requests will increase the longer they wait, automatically prioritizing itself.
  Both aspects contribute to preventing starvation.
\end{denseitemize}

\begin{algorithm}[t]
  \begin{algorithmic}[1]
  \Statex \textbf{Inputs:}
  Number of requests $N$, memory capacity $M$,
  request context length array $l[N]$,
  request QoE gain array $q[N]$, and
  target batch size $B$.
  \Statex \textbf{Output:} Scheduling decision array $x[N]$.

  \ForAll{$i \in [1, N]$}
  \State $p[i] \leftarrow {q[i]}/{l[i]}$ \Comment{Set priority of request $i$}
  \EndFor

  \State $M_{\textrm{current}} \leftarrow N_{\textrm{current}} \leftarrow 0$
  \State Initialize solution array $x[N]$ with all zeros
  \ForAll{$i \in [1, N]$ in descending order of $p[i]$}
      \If{$M_{\textrm{current}}+ l[i] \le M$ \textbf{and} $N_{\textrm{current}} + 1 \le B$}
          \State $x[i] \leftarrow 1$ \Comment{Serve request $i$}
          \State $M_{\textrm{current}} \leftarrow M_{\textrm{current}} + l[i]$
          \State $N_{\textrm{current}} \leftarrow N_{\textrm{current}} + 1$
      \Else
          \State \textbf{break}
      \EndIf
  \EndFor
  \State \Return $x$
  \caption{Priority-based greedy packing for Equation~\ref{eq:scheduling-optimization-problem}}\label{algo:scheduling-greedy}
  \end{algorithmic}
\end{algorithm}

The scheduling algorithm executed at every scheduling quantum for each batch size $B$ is given in Algorithm~\ref{algo:scheduling-greedy}.
In essence, the algorithm sorts requests in descending order of priority and decides to serve the request until the memory capacity or the batch size is reached.
The greedy packing algorithm provides an efficient time complexity of $O(N\log N)$.
On top of this, we apply two optimizations that reduce the number of times the scheduling algorithm must be invoked.

\paragraph{Selective Triggering.}
As QoE is only affected when the system is under load surge, it is not necessary to solve the knapsack problem otherwise.
Therefore, we can selectively trigger Algorithm~\ref{algo:scheduling-greedy} only when we detect that the system is either memory capacity-bound or compute bound.
For the former one, \name monitors the GPU KV cache occupancy and triggers the solver only when occupancy exceeds a high watermark (e.g., 90\%).
For the latter case, \name monitors token generation latency and triggers the solver when it begins to exceed the most minimum token delivery speed requirement of the most stringent request.
 
\paragraph{Batch Size Search Space Pruning.}
In order to further reduce invocations, we reduce the search space of batch size $B$ from $[1, N]$ to $[B_{\min}, B_{\max}]$.
First, there is no point in exploring very large batch sizes that cannot be realized.
Thus, $B_{\max}$ is determined by adding to the batch requests with the shortest context lengths until the total number of tokens in the batch reaches $M$, at which point batch size is the largest possible.
On the other hand, very small batch sizes that can generate tokens faster than the user token consumption speed of \emph{every} request are also suboptimal.
This is because generating tokens that fast does not increase the QoE of served requests, but on the other hand will serve fewer requests, potentially degrading the QoE of requests that are left waiting.
Thus, $B_{\min}$ is set as the largest batch size that generates tokens faster than the most stringent user consumption speed across all requests.

Section~\ref{sec:ablation} shows that this solution can achieve average QoE close to the 3D DP algorithm.

\subsection{Incorporating Preemption Overhead}\label{sec:scheduling-tradeoff}

\begin{figure}[t]
  \centering
  \includegraphics[trim=5 85 120 25,clip,width=0.9\columnwidth]{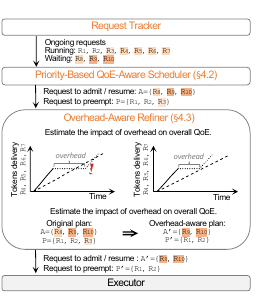} 
  \caption{\name refines the scheduling decision made by the priority-based scheduler so that it does not degrade more QoE than it improves.
    This example scenario is not related to Figure~\ref{fig:solver-overview}.}\label{fig:overhead-toy} 
\end{figure}

In this section, we propose an overhead-aware refiner that improves the scheduling decision made by the priority-based scheduler (\S\ref{sec:scheduling-solution-design}) by considering the impact of preemption overhead on the overall QoE.
This is important because preemptions are not free; they introduce either extra computation or memory movement overhead at the time of preemption and resumption~\cite{shen2024fastswitchoptimizingcontextswitching,vllm-sosp23,infercept-icml24,distserve-osdi24}. 
This overhead can last hundreds of milliseconds to even seconds, and interrupts the whole token generation process.
On the one hand, if too many preemptions occur in a short period, the cumulative overhead can cause significant token generation delays, degrading QoE for all requests. 
On the other hand, if preemption decisions are overly suppressed, the system may miss opportunities to improve overall QoE by serving more urgent requests.

\paragraph{A Motivating Example.}
Figure~\ref{fig:overhead-toy} illustrates how \name's overhead-aware refiner works.
At a high-level, the intuition is that if ongoing requests have enough \emph{slack} before they begin degrading QoE, \name can afford to preempt more requests to maximize overall QoE.
In the toy example, \name's priority-based scheduler (\S\ref{sec:scheduling-solution-design}) decides to admit $\{R_8, R_9, R_{10}\}$ and preempt $\{R_1, R_2, R_{3}\}$, and passes the decision to the overhead-aware refiner.
The overhead-aware refiner then estimates the preemption overhead of each scheduling decision and analyzes its impact on the QoE of all ongoing requests. 
The refiner begins from the highest priority request in the admit list, $R_{10}$, and finds a set of lowest priority requests in the preempt list, $R_1$, that will barely make room for $R_{10}$ when preempted.
With this, \name estimates the total latency overhead of admitting $R_{10}$ and preempting $R_1$.
In this case, \name finds that the QoE gain of admitting $R_{10}$ is higher than the total QoE loss of ongoing requests caused by the admission and preemption overhead.
Thus, $R_{10}$ is added to the final admission list and $R_1$ to the final preemption list.
This process stops when \name finds that the system can no longer tolerate additional overhead without degrading QoE.
Ultimately, the refiner admits only $\{R_9, R_{10}\}$ and preempts $\{R_1, R_2\}$.

\paragraph{Overhead Estimation.}
As seen earlier, \name needs to estimate the overhead of request admission/resumption and preemption.
\name can work with any kind of request preemption mechanism, with the two most frequently available options in existing systems being recomputation and swapping.
The former drops the KV cache of the request upon preemption, incurring no overhead, and then recomputes them on resumption, incurring overhead equivalent to one prefill.
The latter moves the request's KV cache between GPU and CPU memory, which incurs data copying overhead.
Fortunately, overhead is known to be predictable~\cite{clockwork-osdi20,distserve-osdi24,sarathi-serve-osdi24}, so \name offline-profiles the latency overhead of available preemption mechanisms under various context lengths and selects the faster one, and uses that data to estimate overhead for requests.
 
\paragraph{Balancing QoE Gain and Overhead.}
Preemption overhead delays the token generation of every ongoing request, but the scheduling decision made by the priority-based scheduler is unaware of that.
Thus, if the delay from preepmption is excessive, naively implementing the decision as-is may lead to a net degradation in QoE.\@
Therefore, for each admission/resumption decision, the refiner estimates the QoE loss of implementing that decision and prunes those that do not increase more QoE than what it degrades.
On the one hand, QoE gain of admitting/resuming the request was already computed by the priority-aware scheduler as part of the scheduling decision (\S\ref{sec:scheduling-problem-formulation}).
On the other hand, the QoE loss of an ongoing request $i$ can be computed identically with how $Q_{\textrm{wait}, i}$ was estimated, where $\Delta t$ is set to be the preemption overhead.
The total QoE loss of all ongoing requests are then added up.
If the net QoE change is positive, the refiner retains that set of admission/resumption and preemption decisions, and moves on to the next set of requests in priority-order.
Otherwise, it means that the system can no longer tolerate admitting/resuming requests without degrading overall QoE, so the refiner cancels the rest of the admission/resumption and preemption decisions from the priority-based scheduler.

We show in Section~\ref{sec:ablation} that the overhead-aware refiner is critical in maintaining high QoE across requests.
 
\section{Implementation}\label{sec:implementation}

\name mainly consists of the server-side QoE-aware token-level preemptive scheduler and the client-side token pacer.

\paragraph{Request Scheduler.}
\name's scheduling algorithm can work with any LLM serving system that supports continuous batching and at least one preemption mechanism (swapping or recomputation).
As a reference, we implemented \name's scheduler as an alternative scheduling policy in vLLM~\cite{vllm-sosp23}.
The scheduler only manages requests coming into the vLLM instance it is integrated with, assuming that cluster-level load balancing and fault tolerance are done separately.

\begin{figure}[t]
  \centering 
  \begin{subfigure}[c]{0.48\textwidth}
    \centering
    \includegraphics[trim=0 252 0 0,clip,scale=0.35]{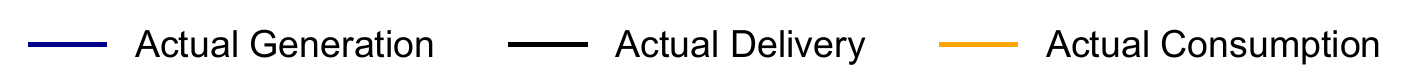} 
  \end{subfigure}

  \begin{subfigure}[c]{0.48\textwidth}
    \centering
    \includegraphics[trim=30 0 0 34,clip,scale=0.40]{new_client_buffer.pdf} 
  \end{subfigure}
  \caption{The client-side token pacer holds excess tokens sent from the server to absorb token generation fluctuations and paces token delivery based on the user's ideal reading speed.}\label{fig:client-buffer}
\end{figure}


\paragraph{Token Pacer.}
The server implements push-based streaming because
(1) tokens generated for a single user are only streamed to that user exactly once, and
(2) the server is typically resource constrained -- particularly so if CPU memory is being used for storing the KV cache of preempted requests -- and therefore prefers to deallocate request state as quickly as possible.
As such, the client receives tokens as soon as they are generated, even if they were generated at a pace that exceeds the user's consumption speed.
Therefore, the client-side \emph{token pacer} temporarily buffers excess tokens and yields them smoothly along the user's Ideal Consumption Timeline.

Figure~\ref{fig:client-buffer} visualizes the token pacer in action.
With an initial burst generation faster than the user's token consumption speed, the pacer withholds excess tokens and paces token delivery, thus growing in size.
Then, the server is aware of the token pacer and the user's QoE parameters, so knowing that this request would have sufficient tokens to deliver to the user for a while, the server preempts the request to serve other requests.
While the request is waiting, the pacer continues to deliver tokens to users at a rate that matches their token consumption speed.
Finally, the server resumes the request at the right timing and starts generating tokens again, and together with the token pacer, perfect QoE was achieved.

\section{Evaluation}\label{sec:eval}

We evaluate \name on a variety of models, hardware configurations (Table~\ref{tab:model-hardware}), and request datasets with varying input and output length characteristics (Table~\ref{tab:dataset}) and find the following:

\begin{denseitemize}
  \item \name effectively handles real-world serving workloads, achieving a QoE of 0.95 or higher for 97\% of served requests, compared to only 75\% for vLLM.\@ \name also reduces peak queue length by 85\%. (\S\ref{sec:e2e-burst})

  \item On synthetic traces with bursts, \name can improve average QoE by up to $4.7\times$ compared to vLLM. 
  Moreover, while preserving high QoE, \name can save up to 61\% GPU resources or alternatively handle up to $2.6\times$ more requests concurrently (\S\ref{sec:e2e-cyclic}).

  \item \name improves QoE by carefully managing the preemptive scheduling overhead, and outperforms baselines under different solver choices, QoE gain estimation time horizons, and request distributions (\S\ref{sec:ablation}). 
\end{denseitemize}

\subsection{Experiment Setup}\label{sec:eval-setup}

\begin{table}[t]
  \centering
  \footnotesize
  \begin{tabular}{llrl}
    \toprule
    \textbf{Model} & \textbf{Architecture} & \textbf{Memory} & \textbf{Hardware} \\
    \midrule
    Phi-3-mini 3.8B~\cite{abdin2024phi}            & Dense, MHA  &   7 GB   & A100$\times$4   \\
    Command R 32B~\cite{cohere_for_ai_2024}        & Dense, GQA  &  61 GB   & A100$\times$8   \\
    Phi-3.5-MoE 16$\times$3.8B~\cite{abdin2024phi} & MoE, GQA    &  80 GB   & A100$\times$8   \\
    Llama 3.1 70B~\cite{llama3-arxiv24}            & Dense, GQA  & 132 GB   & A100$\times$8   \\
    \bottomrule
  \end{tabular}
  \caption{Models and hardware configurations.}\label{tab:model-hardware}
\end{table}

\begin{table}[t]
  \centering
  \footnotesize
  \begin{tabular}{lrrrr}
    \toprule
    \multirow{2}{*}{\textbf{Dataset}} &
    \multicolumn{2}{c}{\textbf{Input Length}} & 
    \multicolumn{2}{c}{\textbf{Output Length}} \\ 
    & Mean & Std. & Mean & Std. \\
    \midrule
    Multi-Round ShareGPT~\cite{sharegpt}  & 3171   & 7943   & 385   & 300    \\
    ArXiv Summarization~\cite{arxiv-sum}  & 17855  & 11401  & 605   & 153    \\
    Coding Challenges~\cite{code-comp}    & 675    & 1552   & 5423  & 21293  \\ 
    \bottomrule
  \end{tabular}
  \caption{Request dataset statistics.}\label{tab:dataset}
\end{table}
   
\textbf{Models, Hardware and Requests.}
We evaluate \name on models with diverse architectures (Dense vs. Mixture-of-Experts, Multi-Head Attention vs. Grouped-Query Attention) and request input/output datasets with varying sequence lengths.
We leverage both real-world request traces from BurstGPT~\cite{burstgpt-arxiv24} and synthetic bursty traces designed to conduct controlled experiments as detailed in Section~\ref{sec:e2e-burst} and Section~\ref{sec:e2e-cyclic} respectively.
We deployed models with tensor parallelism on NVIDIA A100 SXM4 40GB GPUs in one AWS p4d.24xlarge instance. 
See Tables~\ref{tab:model-hardware} and~\ref{tab:dataset} for full details.

\paragraph{QoE Parameters.}
In real-world scenarios, QoE parameters (target TTFT and token consumption speed) of users will vary widely.
For each request, we set the target TTFT to be $\max(\text{input length} / 5000,~1)$ seconds, where 5000 tokens/second is the prefill throughput of our hardware setup.
This assumes that requests with longer input ought to expect proportionally longer TTFT.\@
On the other hand, we use the human reading speed distribution in Figure~\ref{fig:expected-tds-age} to assign user token reading speeds to requests.
In real applications, this should be set depending on the application's specific use case.
Furthermore, API price tiering can also be implemented by providing a higher token consumption speed for higher-priced tokens, allowing users to select the tier suitable for their downstream text streaming application.

\paragraph{Baselines.} 
We compare with vLLM~\cite{vllm-sosp23} (v0.6.1) and Sarathi-Serve~\cite{sarathi-serve-osdi24}, all of which adopts the FCFS scheduling policy.
Additionally, we configure vLLM to use a custom scheduling algorithm, Least QoE Slack First (LQSF), which prioritizes requests most at risk of QoE degradation based on the QoE gain we defined.

\paragraph{Metrics.}
We report the following metrics:
\begin{denseitemize}
  \looseness=-1
  \item \textbf{Average QoE}: The QoE value of requests when they finish, averaged across all requests.


  \item \textbf{Resource savings} (\%): How much fewer GPUs \name needs to maintain an average QoE of 0.95.
\end{denseitemize}

\subsection{End-to-End Improvements on Real-World Traces}\label{sec:e2e-burst}

\begin{figure}[t]
  \centering 
  \begin{subfigure}[b]{0.48\textwidth}
    \centering
    \includegraphics[trim=50 315 355 0,clip,scale=0.33]{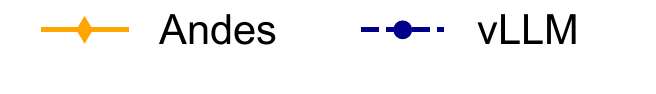} 
  \end{subfigure}
  \begin{subfigure}[b]{0.16\textwidth}
    \centering
    \includegraphics[trim=0 10 0 0,clip,width=0.99\linewidth]{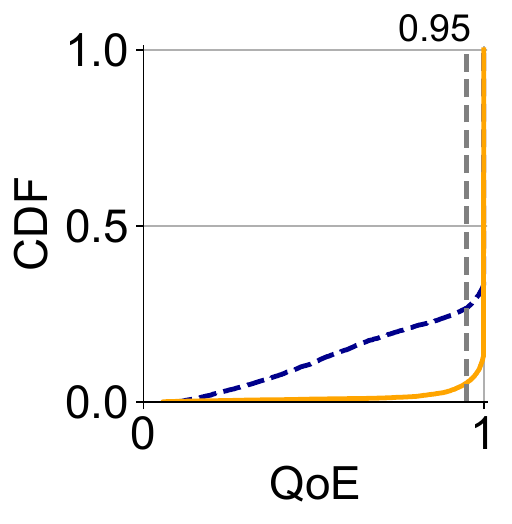}
    \caption{QoE}\label{fig:breakdown-qoe}
  \end{subfigure}%
  \begin{subfigure}[b]{0.16\textwidth}
    \centering
    \includegraphics[trim=0 10 0 0,clip,width=0.96\linewidth]{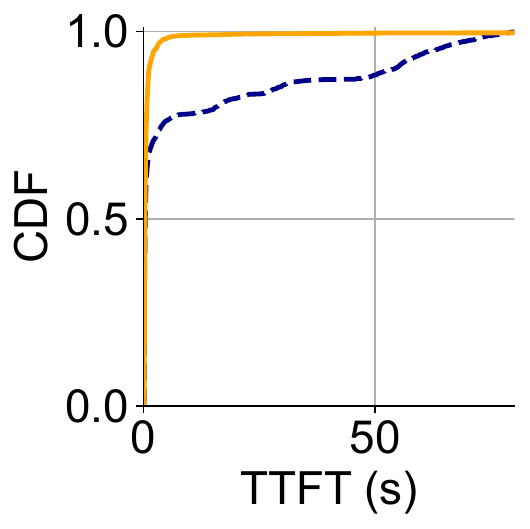}
    \caption{TTFT}\label{fig:breakdown-ttft}
  \end{subfigure}%
  \begin{subfigure}[b]{0.16\textwidth}
    \centering
    \includegraphics[trim=0 10 0 0,clip,width=0.99\linewidth]{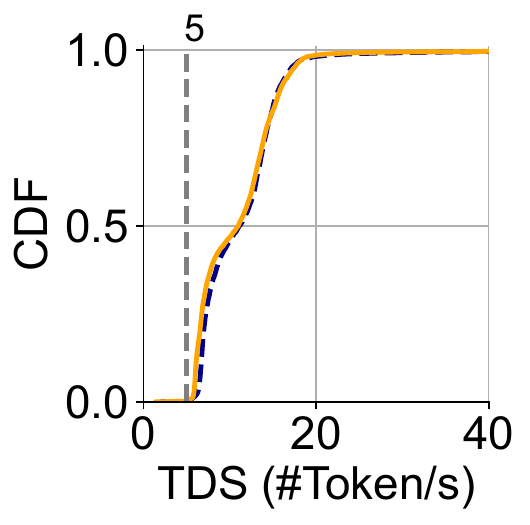 }
    \caption{TDS}\label{fig:breakdown-tds}
  \end{subfigure}%
  \caption{QoE, TTFT, and TDS CDFs of requests in BurstGPT. }\label{fig:breakdown} 
\end{figure}

\begin{figure}[t]
  \begin{subfigure}[b]{0.47\textwidth}
    \centering 
    \includegraphics[trim=0 290 0 10,clip,width=0.6\linewidth]{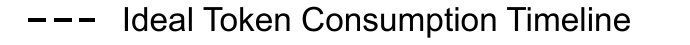 } 
  \end{subfigure}

  \begin{subfigure}[b]{0.24\textwidth}
    \includegraphics[trim=0 25 0 0,clip,width=0.95\linewidth]{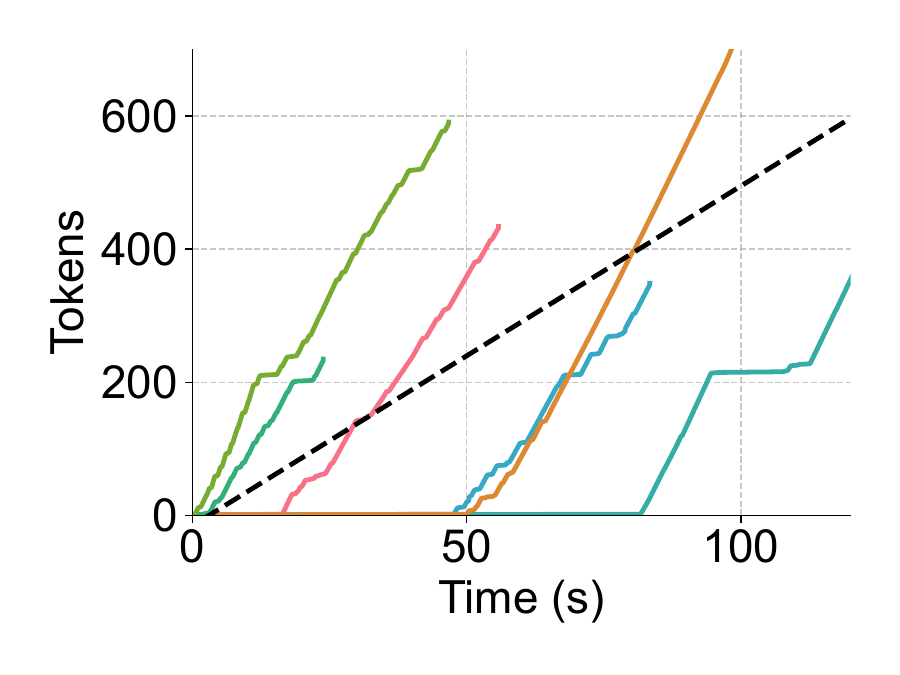}
    \caption{vLLM}
  \end{subfigure}%
  \begin{subfigure}[b]{0.24\textwidth}
    \includegraphics[trim=0 25 0 0,clip,width=0.95\linewidth]{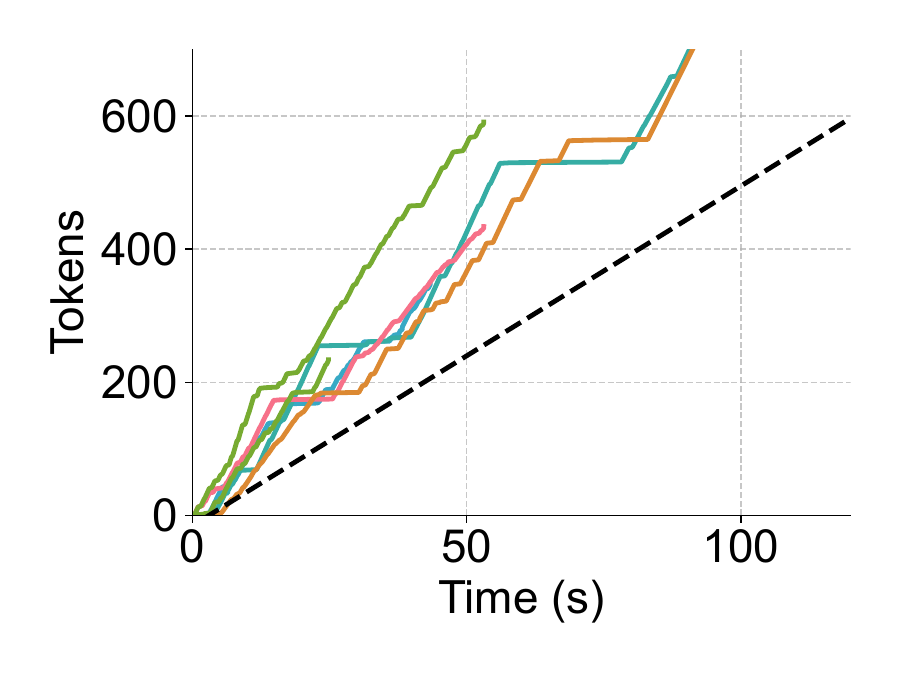 }
    \caption{\name}
  \end{subfigure}%
  \caption{
    The Actual Delivery Timeline of select requests, with request submission time shifted to zero.
    Requests served by \name, as opposed to vLLM, are well above the Ideal Consumption Timeline.}\label{fig:tdt} 
\end{figure}

\begin{figure}[t]
  \centering 
  \includegraphics[trim=0 15 0 0,clip,width=0.49\textwidth]{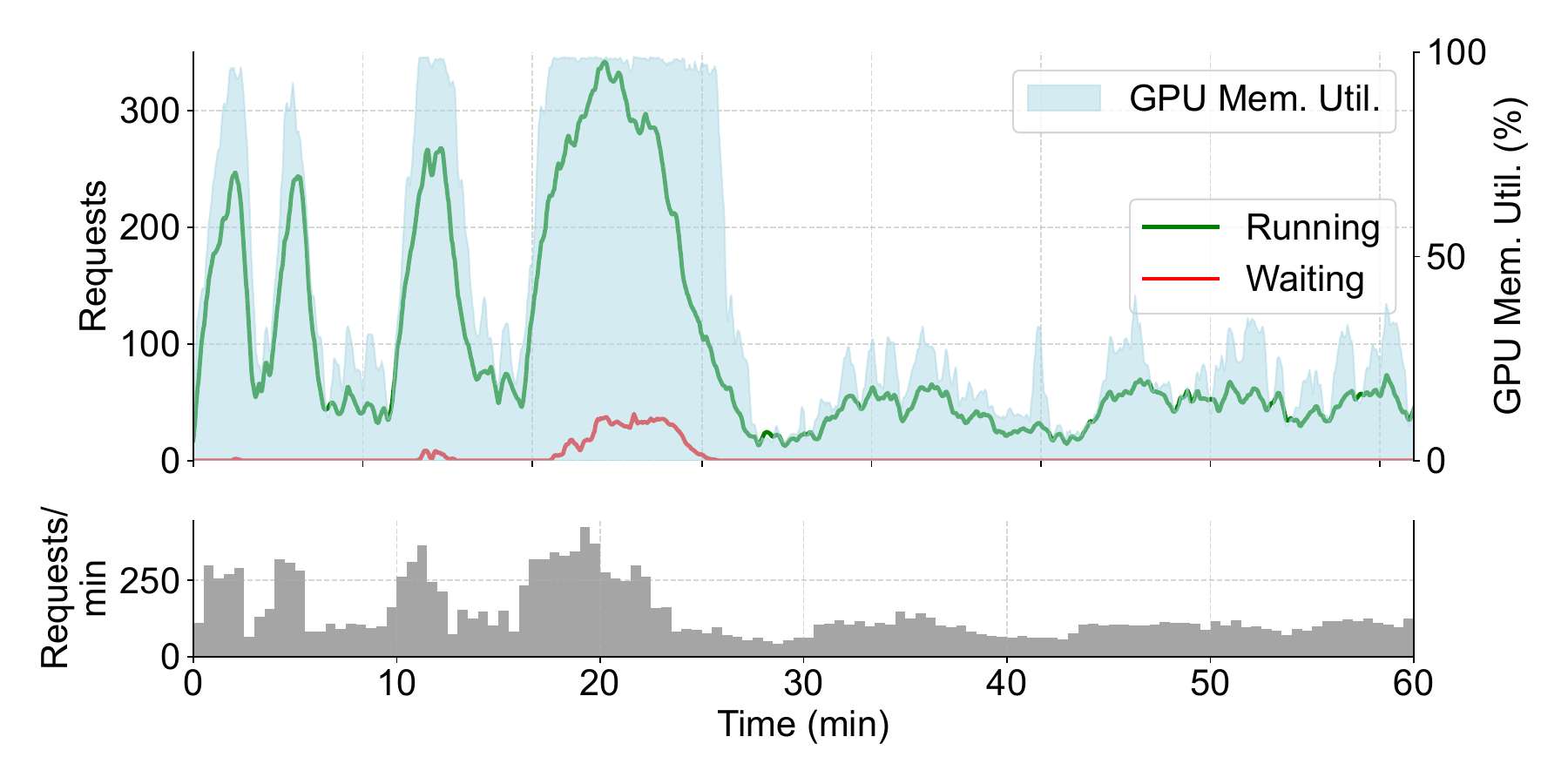}     
  \caption{
    \name serving requests from BurstGPT.\@
    vLLM serving the same trace is shown in Figure~\ref{fig:burst-fcfs}.
    Queue length is significantly reduced under load surges, and GPU memory utilization is higher.}\label{fig:burst-qoe}
\end{figure}

We replay a one-hour slice of BurstGPT~\cite{burstgpt-arxiv24}, a request trace captured from real-world LLM services, with the Multi-Round ShareGPT input/output dataset in order to show the end-to-end QoE improvement of \name.

\paragraph{QoE Improvement.}
We report the CDF of QoE, TTFT, and TDS across all requests in Figure~\ref{fig:breakdown}.
Overall, compared with vLLM using FCFS, \name improves average QoE from 0.88 to 0.99 and reduces average TTFT from 10.5s to 1.8s.
Notably, 97\% of requests served by \name achieve a QoE of 0.95 or higher, compared to only 75\% under vLLM.\@

QoE improvement without additional resources means that \name can serve more requests concurrently while maintaining high QoE levels, or conversely, significantly reduce the amount of GPUs needed to maintain the same level of QoE, directly leading to cost savings.
\name achieves this while reducing the average TDS of requests only by a marginal amount -- from 11.2 tokens/s to 10.9 tokens/s -- which is enough to satisfy most user token consumption speeds, showing that \name keeps the overhead of preemptive scheduling well under control.

\paragraph{Token Generation Timeline Improvements.}
We select a set of requests from the trace and visualize how they were served differently by vLLM and \name in Figure~\ref{fig:tdt}.
We can observe that \name is able to generate every token before the ideal consumption time through QoE-aware token-level preemptive scheduling, while vLLM suffers from head-of-line blocking and degrades the user experience of many requests.
With the help of token pacer, the request's token delivery timeline can perfectly align with the ideal consumption timeline.

\paragraph{Significant Queue Length Reduction.}
To examine the system's real-time behavior, we visualize the state of \name while serving requests in Figure~\ref{fig:burst-qoe}.
Visualization of vLLM serving the same trace can be found in Figure~\ref{fig:burst-fcfs}.
We observe that \name can reduce peak waiting queue length during serving by a significant 85\% through token-level preemptive request scheduling.
Furthermore, unlike vLLM, a large portion of the waiting requests in \name are those that have been preempted by the scheduler after generating sufficient tokens for their users, explaining the high average QoE achieved by \name.

%

\subsection{Performance Under Controlled Burstiness}\label{sec:e2e-cyclic}

\begin{figure}[t]
  \centering 
  \includegraphics[trim=0 0 0 0,clip,width=0.35\textwidth]{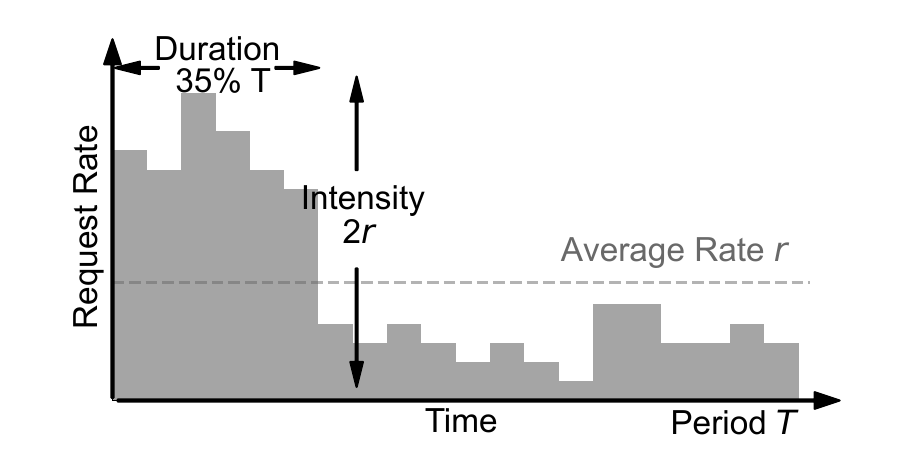}     
  \caption{One cycle of the cyclic burst load pattern.}\label{fig:cyclic-burst}
\end{figure}

\begin{figure}[t]
  \centering
  \begin{subfigure}[b]{0.5\textwidth}
   \centering
   \includegraphics[trim=50 325 10 0,clip,scale=0.33]{e2e-legend.pdf}
\end{subfigure}

\begin{subfigure}[b]{0.15\textwidth}
  \centering
  \includegraphics[trim=0 0 0 0,clip,scale=0.33]{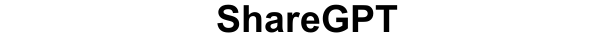} 
\end{subfigure}%
\begin{subfigure}[b]{0.15\textwidth}
  \centering
  \includegraphics[trim=0 0 0 0,clip,scale=0.33]{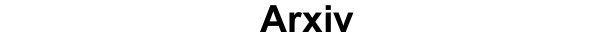} 
\end{subfigure}%
  \begin{subfigure}[b]{0.15\textwidth}
    \centering
    \includegraphics[trim=0 0 0 0,clip,scale=0.33]{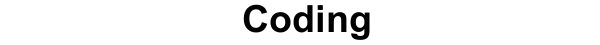} 
  \end{subfigure}%

\begin{subfigure}[b]{0.03\textwidth}
  \includegraphics[trim=0 0 0 0,clip,scale=0.33]{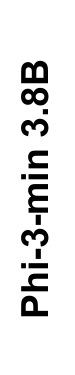}
\end{subfigure}%
   \begin{subfigure}[b]{0.16\textwidth}
       \includegraphics[trim=0 0 0 0,clip,scale=0.25]{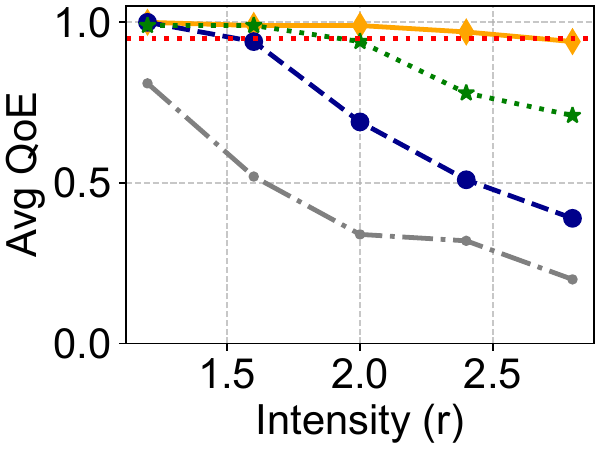}
   \end{subfigure}%
   \begin{subfigure}[b]{0.15\textwidth}
     \includegraphics[trim=26 0 0 0,clip,scale=0.25]{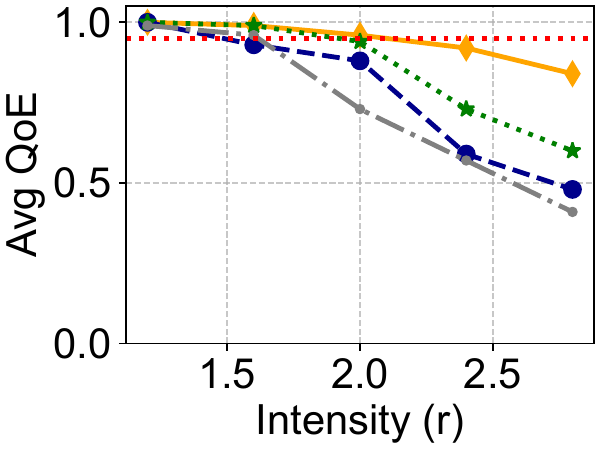}
     \end{subfigure}%
       \begin{subfigure}[b]{0.15\textwidth}
         \includegraphics[trim=26 0 0 0,clip,scale=0.25]{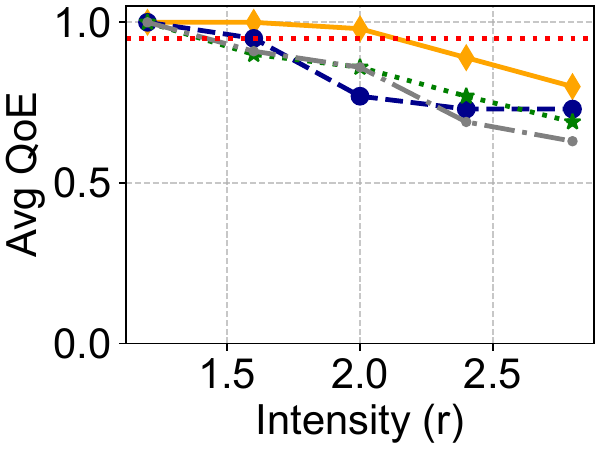}
       \end{subfigure}%

\begin{subfigure}[b]{0.03\textwidth}
  \includegraphics[trim=0 0 0 0,clip,scale=0.33]{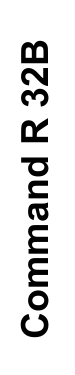} 
\end{subfigure}%
   \begin{subfigure}[b]{0.16\textwidth}
     \includegraphics[trim=0 0 0 0,clip,scale=0.25]{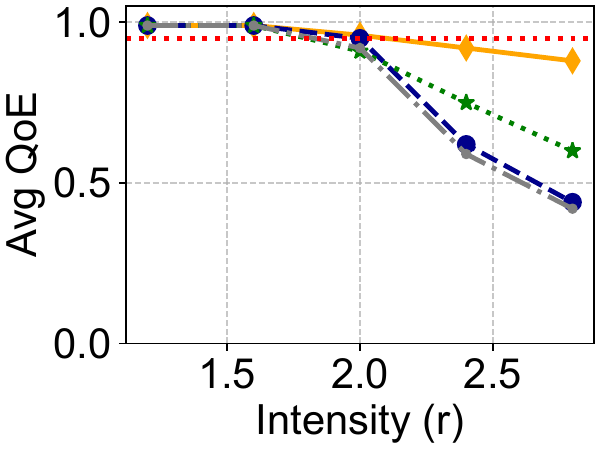}
 \end{subfigure}%
 \begin{subfigure}[b]{0.15\textwidth}
   \includegraphics[trim=26 0 0 0,clip,scale=0.25]{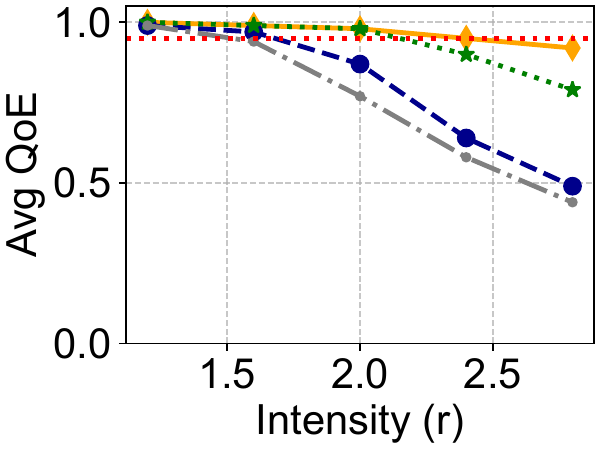}
   \end{subfigure}%
     \begin{subfigure}[b]{0.15\textwidth}
       \includegraphics[trim=26 0 0 0,clip,scale=0.25]{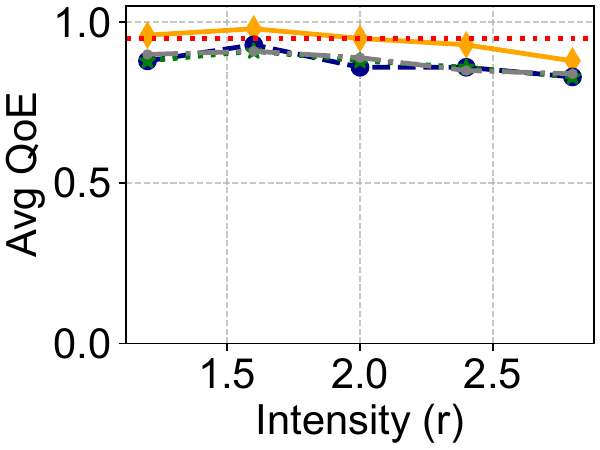}
     \end{subfigure}%

 
\begin{subfigure}[b]{0.03\textwidth}
  \includegraphics[trim=0 0 0 0,clip,scale=0.33]{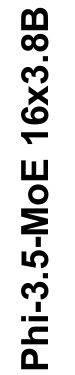}
\end{subfigure}%
  \begin{subfigure}[b]{0.16\textwidth}
     \includegraphics[trim=0 0 0 0,clip,scale=0.25]{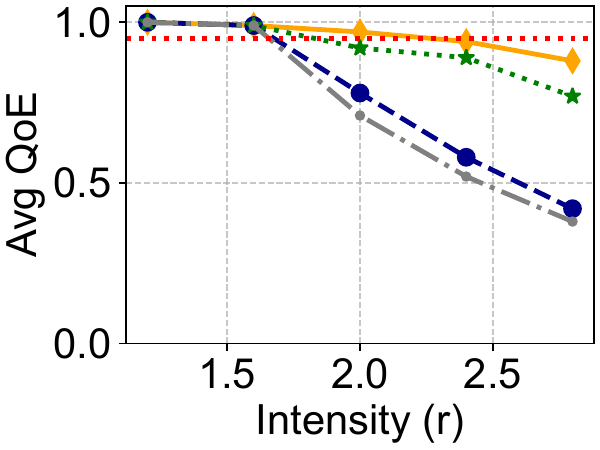}
 \end{subfigure}%
 \begin{subfigure}[b]{0.15\textwidth}
   \includegraphics[trim=26 0 0 0,clip,scale=0.25]{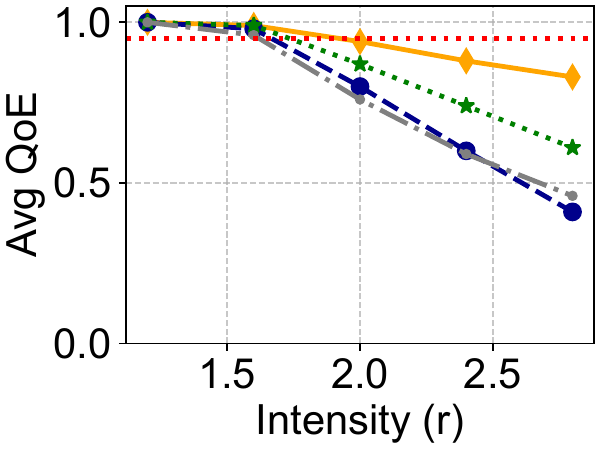}
   \end{subfigure}%
     \begin{subfigure}[b]{0.15\textwidth}
       \includegraphics[trim=26 0 0 0,clip,scale=0.25]{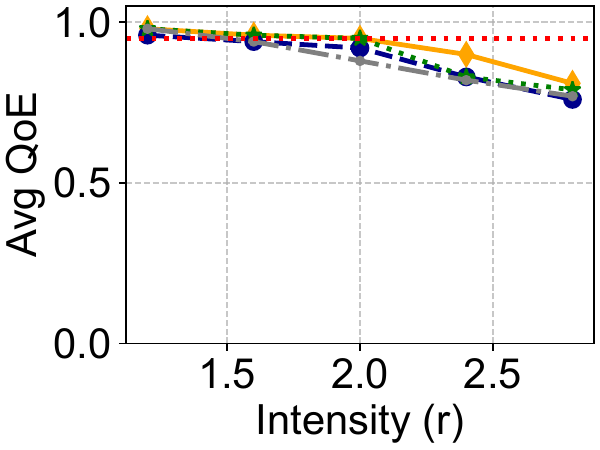}
     \end{subfigure}%

\begin{subfigure}[b]{0.03\textwidth}
  \includegraphics[trim=0 0 0 0,clip,scale=0.33]{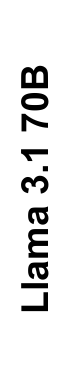}
\end{subfigure}%
   \begin{subfigure}[b]{0.16\textwidth}
     \includegraphics[trim=0 0 0 0,clip,scale=0.25]{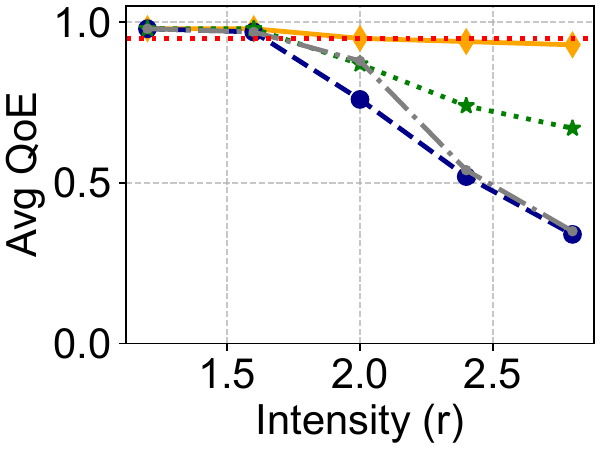}
 \end{subfigure}%
 \begin{subfigure}[b]{0.15\textwidth}
   \includegraphics[trim=26 0 0 0,clip,scale=0.25]{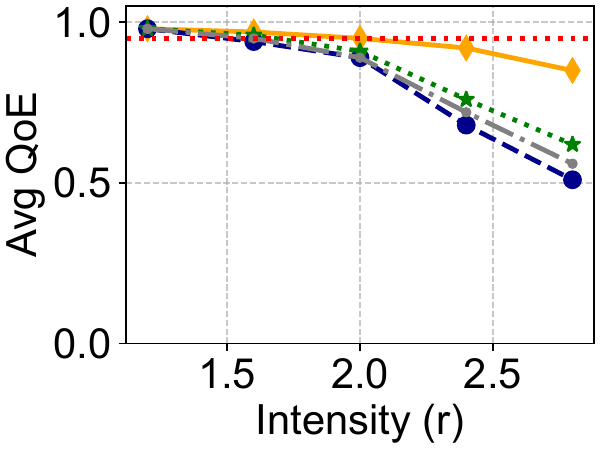}
   \end{subfigure}%
     \begin{subfigure}[b]{0.15\textwidth}
       \includegraphics[trim=26 0 0 0,clip,scale=0.25]{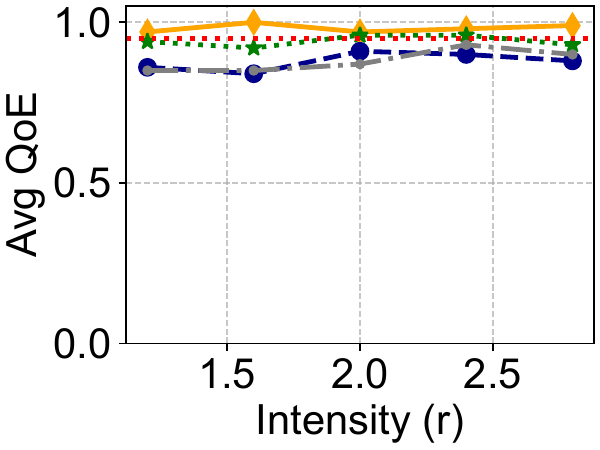}
     \end{subfigure}%

     \caption{Average QoE while varying burst intensity.}
     \label{fig:e2e-intensity} 
\end{figure}

\begin{figure}[t]
  \centering
  \begin{subfigure}[b]{0.5\textwidth}
   \centering
   \includegraphics[trim=50 325 10 0,clip,scale=0.33]{e2e-legend.pdf}
  \end{subfigure}

\begin{subfigure}[b]{0.15\textwidth}
  \centering
  \includegraphics[trim=0 0 0 0,clip,scale=0.33]{ShareGPT.pdf} 
\end{subfigure}%
\begin{subfigure}[b]{0.15\textwidth}
  \centering
  \includegraphics[trim=0 0 0 0,clip,scale=0.33]{Arxiv.pdf} 
\end{subfigure}%
  \begin{subfigure}[b]{0.15\textwidth}
    \centering
    \includegraphics[trim=0 0 0 0,clip,scale=0.33]{Coding.pdf} 
  \end{subfigure}%

\begin{subfigure}[b]{0.03\textwidth}
  \includegraphics[trim=0 0 0 0,clip,scale=0.33]{Phi-3-min_3.8B.pdf} 
\end{subfigure}%
   \begin{subfigure}[b]{0.16\textwidth}
       \includegraphics[trim=0 0 0 0,clip,scale=0.25]{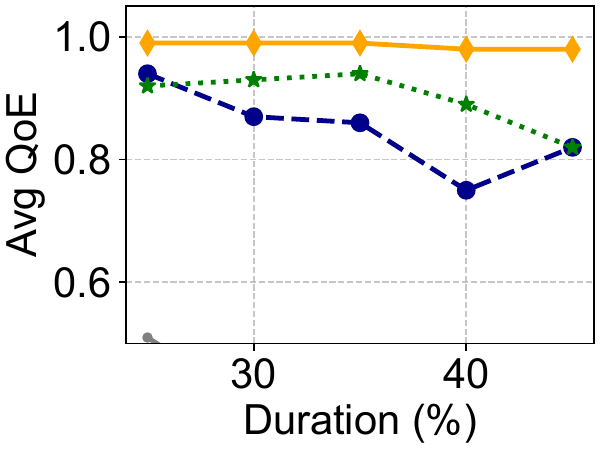}
   \end{subfigure}%
   \begin{subfigure}[b]{0.15\textwidth}
     \includegraphics[trim=26 0 0 0,clip,scale=0.25]{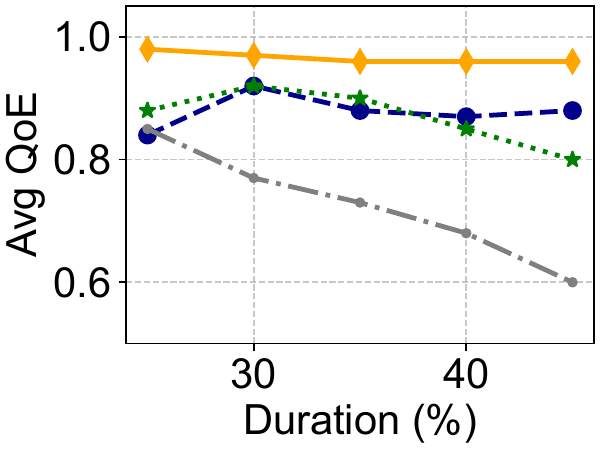}
     \end{subfigure}%
       \begin{subfigure}[b]{0.15\textwidth}
         \includegraphics[trim=26 0 0 0,clip,scale=0.25]{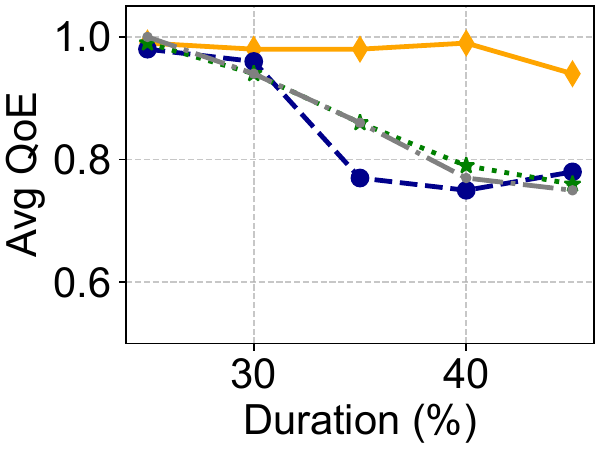}
       \end{subfigure}%

\begin{subfigure}[b]{0.03\textwidth}
  \includegraphics[trim=0 0 0 0,clip,scale=0.33]{Command_R_32B.pdf}
\end{subfigure}%
   \begin{subfigure}[b]{0.16\textwidth}
     \includegraphics[trim=0 0 0 0,clip,scale=0.25]{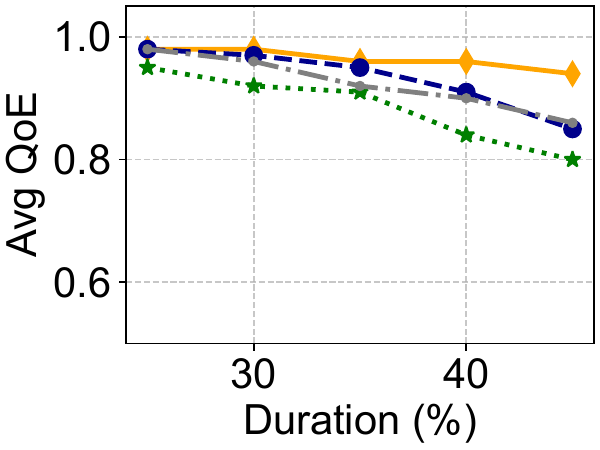}
 \end{subfigure}%
 \begin{subfigure}[b]{0.15\textwidth}
   \includegraphics[trim=26 0 0 0,clip,scale=0.25]{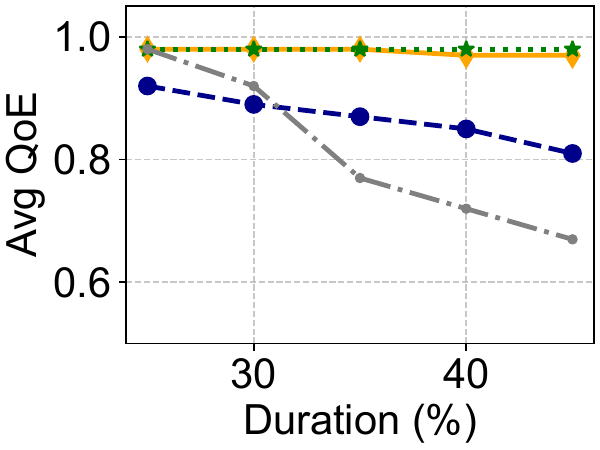}
   \end{subfigure}%
     \begin{subfigure}[b]{0.15\textwidth}
       \includegraphics[trim=26 0 0 0,clip,scale=0.25]{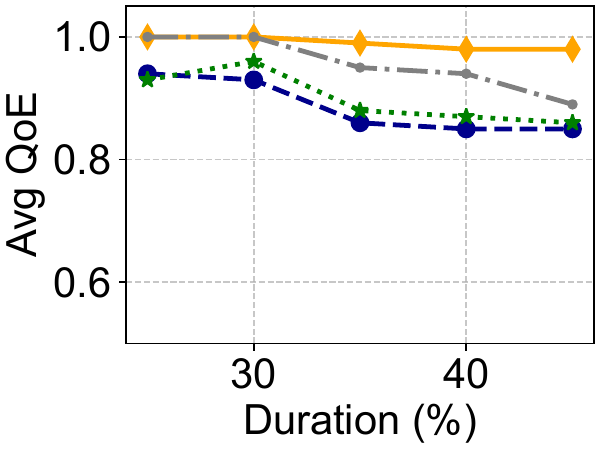}
     \end{subfigure}%

 
\begin{subfigure}[b]{0.03\textwidth}
  \includegraphics[trim=0 0 0 0,clip,scale=0.33]{Phi-3.5-MoE_16x3.8B.pdf}
\end{subfigure}%
  \begin{subfigure}[b]{0.16\textwidth}
     \includegraphics[trim=0 0 0 0,clip,scale=0.25]{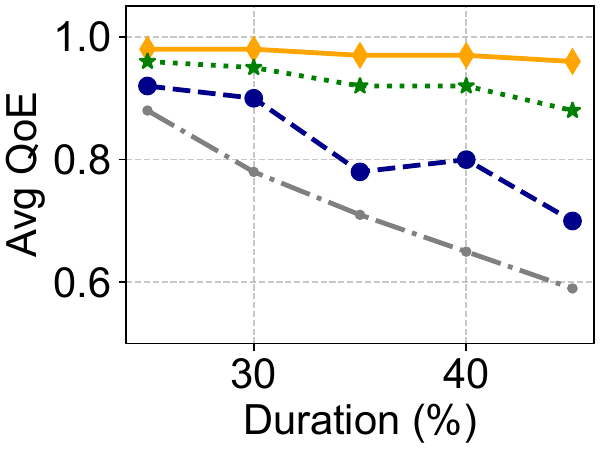}
 \end{subfigure}%
 \begin{subfigure}[b]{0.15\textwidth}
   \includegraphics[trim=26 0 0 0,clip,scale=0.25]{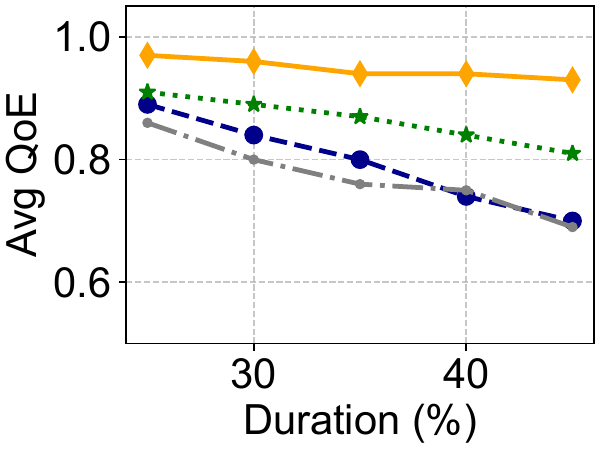}
   \end{subfigure}%
     \begin{subfigure}[b]{0.15\textwidth}
       \includegraphics[trim=26 0 0 0,clip,scale=0.25]{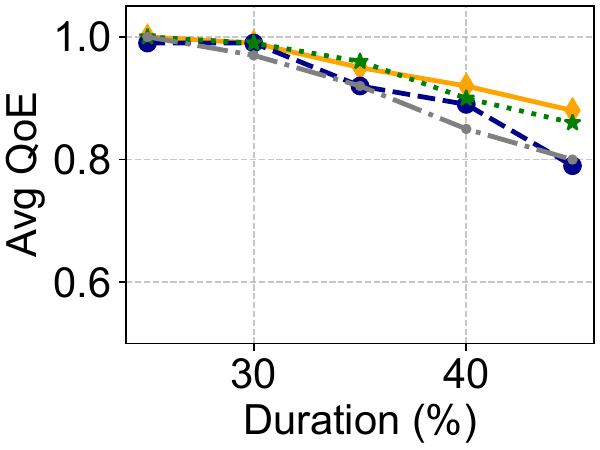}
     \end{subfigure}%

\begin{subfigure}[b]{0.03\textwidth}
  \includegraphics[trim=0 0 0 0,clip,scale=0.33]{Llama_3.1_70B.pdf}
\end{subfigure}%
   \begin{subfigure}[b]{0.16\textwidth}
     \includegraphics[trim=0 0 0 0,clip,scale=0.25]{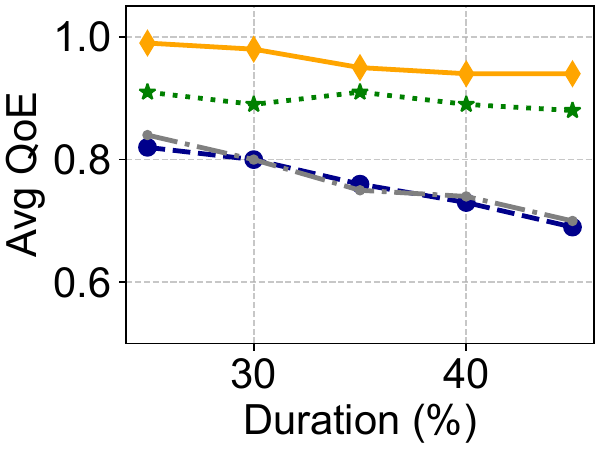}
 \end{subfigure}%
 \begin{subfigure}[b]{0.15\textwidth}
   \includegraphics[trim=26 0 0 0,clip,scale=0.25]{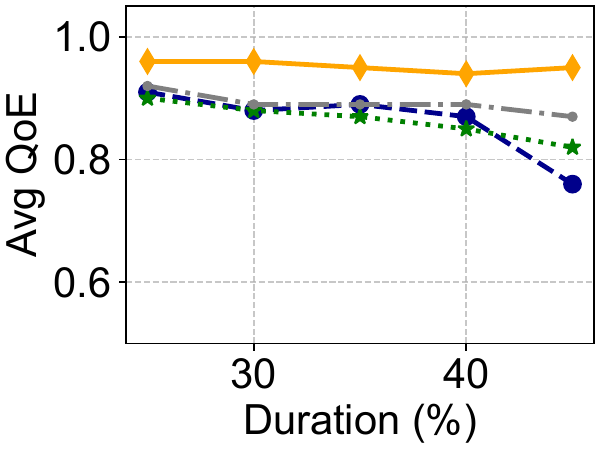}
   \end{subfigure}%
     \begin{subfigure}[b]{0.15\textwidth}
       \includegraphics[trim=26 0 0 0,clip,scale=0.25]{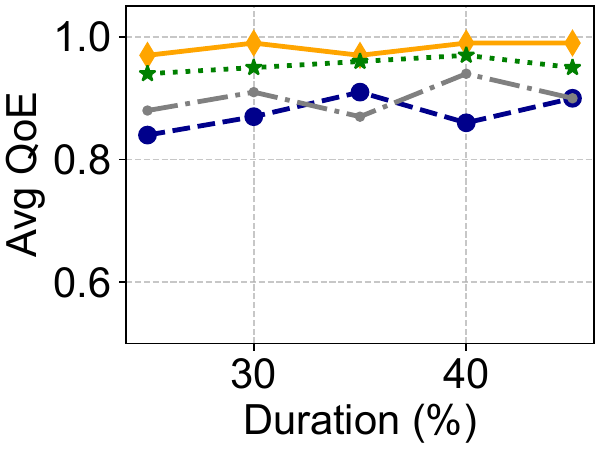}
     \end{subfigure}%

  \caption{Average QoE while varying burst duration.}
     \label{fig:e2e-duration} 
\end{figure}

We have seen that \name can effectively handle requests from real-world traces.
In this section, in order to quantify the QoE improvement and resource savings of \name compared to baselines, we evaluate \name on a set of synthetic traces that contain request bursts that resemble those in BurstGPT.\@

More specifically, BurstGPT shows an average of three burst periods per hour, with each burst sustaining an average duration of 7 minutes and showing on average 2$\times$ higher request rates during bursts.
Following this, we introduce the \emph{Cyclic Burst Load Pattern}, which resembles burst patterns in BurstGPT.\@
Figure~\ref{fig:cyclic-burst} illustrates one cycle of burst in the cyclic burst load pattern.
Bursts are characterized with two parameters: \emph{intensity} (ratio of request rate from burst to the entire trace) and \emph{duration} (percentage of time the burst takes up in the whole trace).
Request arrivals in both burst and non-burst phases follow the Poisson process, and burst/non-burst periods alternate in a cyclic pattern.

Following statistics from BurstGPT, we set the default burst intensity to 2 and the burst duration to 35\%. 
In our experiments, we adjust either burst intensity or burst duration while leaving the other at its default value.
The average request rate of the whole cycle is set to be the serving system's throughput without any burstiness, preventing the system's queue from growing indefinitely over time.
This also allows us to evaluate systems on one cycle of the trace, as most requests will have been handled by the end of one cycle.

\paragraph{Improvements Under Varying Burst Intensity.}
We report the average QoE under different burst intensities in Figure~\ref{fig:e2e-intensity}.
As burst intensity increases, \name continues to maintain high average QoE, achieving up to $4.7\times$ QoE improvement.
We draw a horizontal line on average QoE 0.95 to find the maximum burst intensity each system can sustain while keeping that level of QoE.\@
It can be seen that, compared to vLLM, \name can handle up to $2.6\times$ more burst intensity than vLLM while maintaining average QoE, or alternatively, save up to $61\%$ GPU resources. 

The gap is particularly large with baselines that use FCFS scheduling, as they are susceptible to head-of-line blocking during burst periods and degrades QoE.\@
In addition, Sarathi-Serve performs poorly under some cases because chunked prefill interferes with the prefill stage and increases TTFT.\@ 
The Least QoE Slack First (LQSF) scheduling algorithm shows a slight QoE improvement, as it prioritizes requests that are most at risk of QoE degradation, based on thier QoE gain.
However, as it fails to account for the resource usage of requests, causing requests with more resource usage to starve other requests under limited GPU resources, it still falls short of \name's performance.

\paragraph{Improvements Under Varying Burst Duration.}
We also report the average QoE under different burst durations in Figure~\ref{fig:e2e-duration}.
\name consistently provides higher average QoE across all models and request datasets; overall, \name delivers up to $3.5\times$ higher average QoE compared to vLLM.\@ 

\paragraph{How Much Potential Gain was Realized?}
As derived in Section~\ref{sec:motivation-opportunities}, \name can serve $2.3\times$ more requests under ideal circumstances for the case of Phi-3.5-MoE 16$\times$3.8B with the Multi-Round ShareGPT request dataset.
In practice, as shown in Figure~\ref{fig:e2e-intensity}, \name achieves a $1.4\times$ improvement, realizing approximately $60\%$ of the ideal gain under this setup.
This gap is due to scheduling overhead, request prefill, and request preempt/resume overhead, which limit the full utilization of the system's available slack.

\subsection{Sensitivity Analysis and Ablation Studies}\label{sec:ablation}

We evaluate the robustness of \name under diverse settings and system configurations.
We report our results with the Llama 3.1 70B model on the Multi-Round ShareGPT dataset and default cyclic burst load pattern as we observed similar trends in different setups.
 
\begin{figure}[t]
  \begin{subfigure}[b]{0.47\textwidth}
    \centering
    \includegraphics[trim=0 210 0 0,clip,width=\linewidth]{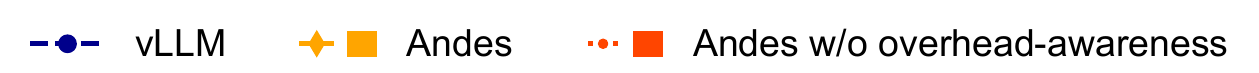}
  \end{subfigure}

  \begin{subfigure}[b]{0.23\textwidth}
    \centering
    \includegraphics[trim=0 0 0 0,clip,width=0.9\linewidth]{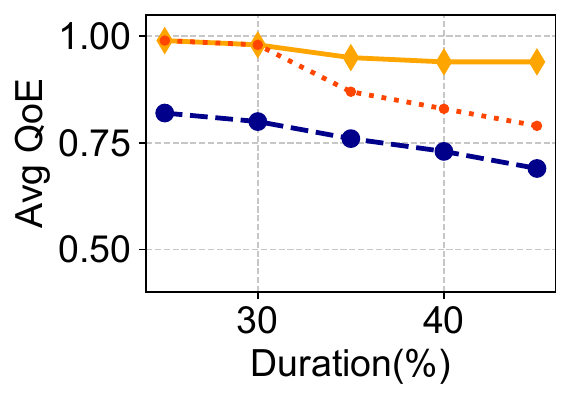}
  \end{subfigure}
  \begin{subfigure}[b]{0.23\textwidth}
    \centering
    \includegraphics[width=0.9\linewidth]{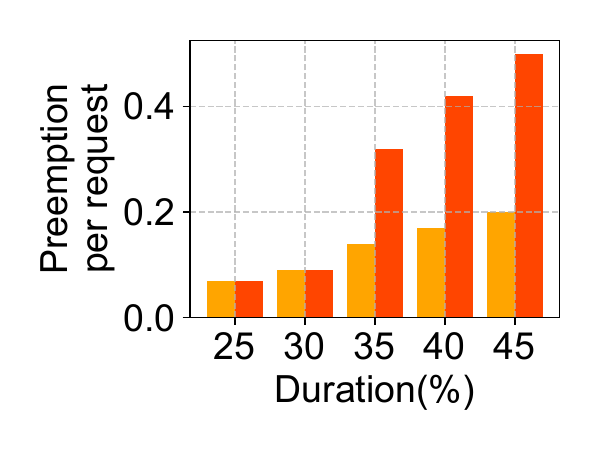}
  \end{subfigure} 
  \caption{Overhead-aware refiner is critical to optimize QoE.}\label{fig:overhead-ablation} 
\end{figure}

\paragraph{Overhead-Aware Refiner is Indispensable.}
To assess the effectiveness of overhead-aware refiner (\S\ref{sec:scheduling-tradeoff}), we conduct an ablation study comparing \name's overhead-aware refiner to a baseline that ignores preemption overheads.
Figure~\ref{fig:overhead-ablation} shows the average QoE and the average number of preemptions per request under different burst durations.
\name without the overhead-aware refiner makes scheduling decisions that are unaware of preemption overhead.
As such, it incurs excessive preemptions with the increase of burst duration, which delays token generation for ongoing requests and thus significantly degrading QoE. 
In contrast, \name (with the overhead-aware refiner) strikes a balance between QoE improvement and scheduling costs, achieving consistently higher QoE under different burstiness setups.

\begin{figure}[t]
  \begin{subfigure}[b]{0.47\textwidth}
    \centering 
    \includegraphics[trim=0 220 0 0,clip,width=0.90\linewidth]{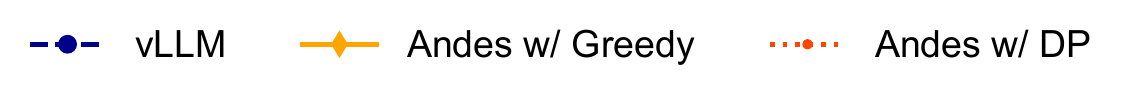}
  \end{subfigure}

  \begin{subfigure}[b]{0.23\textwidth}
    \centering 
    \includegraphics[width=0.9\linewidth]{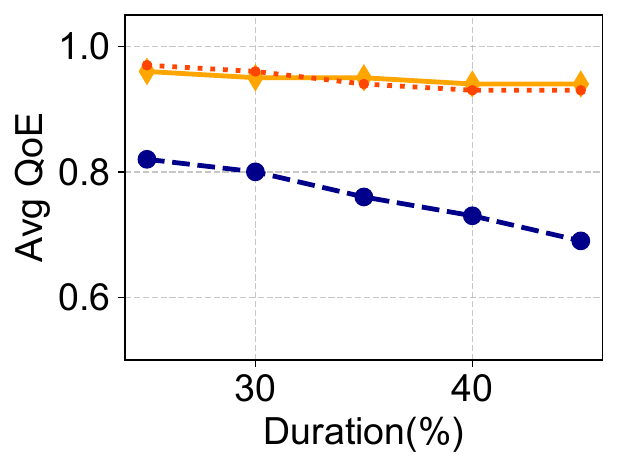}
  \end{subfigure}
  \begin{subfigure}[b]{0.23\textwidth}
    \centering 
    \includegraphics[width=0.9\linewidth]{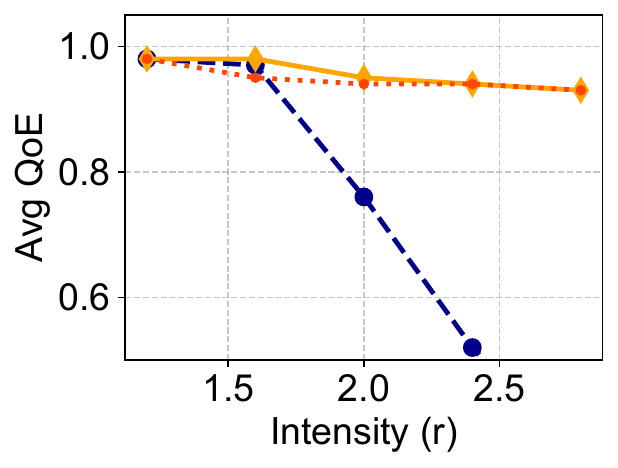}
  \end{subfigure}
  \caption{Different knapsack solver.}\label{fig:knapsack-solver} 
\end{figure}

\paragraph{Different Knapsack Solvers.}
We compare \name's greedy knapsack solver with the optimal-but-slow 3D Dynamic Programming solver.
Figure~\ref{fig:knapsack-solver} shows that the greedy solver achieves slightly better average QoE under longer burst durations or higher burst intensities.
\name outperforms the exact 3D DP solver because its greedy solver is more suitable for real-time deicision making, being $\sim 20\times$ faster while still delivering high-quality approximate solutions.

\begin{figure}[t]
  \begin{minipage}[b]{0.48\textwidth}
    \begin{figure}[H]
      \centering
      \includegraphics[trim=50 325 10 0,clip,scale=0.33]{e2e-legend.pdf}
    \end{figure}
  \end{minipage}

  \begin{minipage}[b]{0.23\textwidth}
    \begin{figure}[H]
      \centering
      \includegraphics[width=0.9\linewidth]{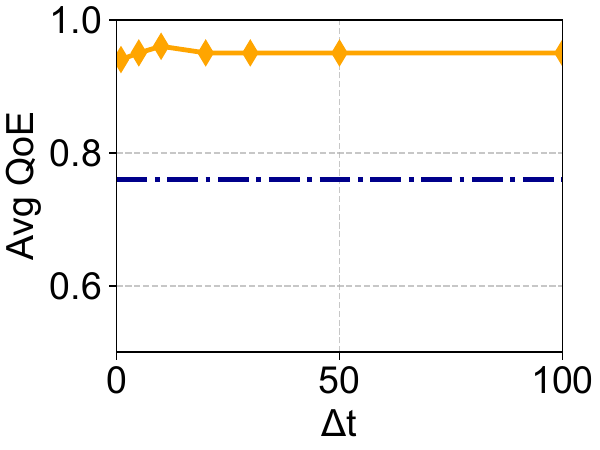}
      \caption{Varying $\Delta t$.}\label{fig:delta-t}
    \end{figure}
  \end{minipage}
  \begin{minipage}[b]{0.23\textwidth}
    \begin{figure}[H]
      \centering
      \includegraphics[width=0.9\linewidth]{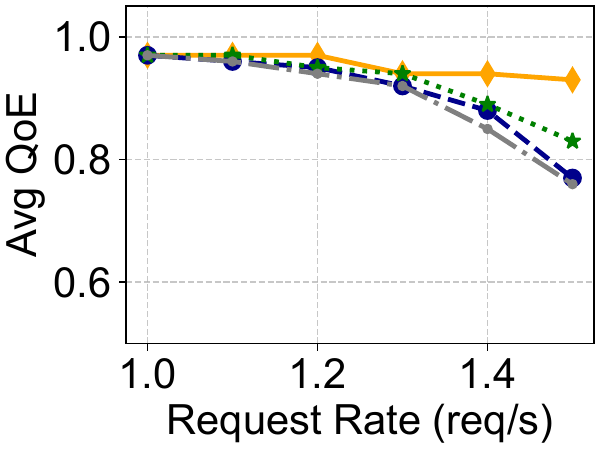}
      \caption{Poisson arrival.}\label{fig:poisson}
    \end{figure}
  \end{minipage}
\end{figure}

\paragraph{QoE Gain Estimation Time Horizon.}
We evaluate how varying $\Delta t$, the time horizon over which a request's QoE gain is estimated, influences average QoE.\@
Figure~\ref{fig:delta-t} shows that the average QoE remains roughly consistent for various $\Delta t$ values and significantly outperforms the baselines.
The best value of $\Delta t$ depends on both the model and request input/output distribution, requiring fine-tuning prior to deployment.

\paragraph{Different Arrival Distributions.}
In our main evaluation, we use the real-world BurstGPT trace and synthetic traces that resemble BurstGPT's load surges.
For completeness, we present results from running \name and baselines on a Poisson arrival trace.
Figure~\ref{fig:poisson} shows the average QoE of systems under varying Poisson arrival rates over a duration of 20 minutes.
\name still consistently delivers higher average QoE compared to baselines, particularly under high request rates.

\section{Related Work}

\paragraph{LLM Inference Serving.}
Orca~\cite{orca-osdi22} introduced iteration-level batching to enhance the throughput of LLM inference, followed by vLLM~\cite{vllm-sosp23} developing PagedAttention to optimize memory usage.
Splitwise~\cite{splitwise-isca24}, DistServe~\cite{distserve-osdi24},
Sarathi-Serve~\cite{sarathi-serve-osdi24}, and LoongServe~\cite{loongserve-sosp24} optimize prefill and decode computations, but keeps the FCFS request scheduler.
VTC~\cite{sheng2024fairness-osdi24} develops a non-preemptive request scheduler based on fairness, Llumnix~\cite{llumnix-osdi24} proposes cluster-wide request live migration for various types of load balancing, and CacheGen~\cite{cachegen-sigcomm24} incrementally streams reusable KV cache entries to LLM server instances.
To the best of our knowledge, \name is the first to identify the insufficiency of existing optimization metrics for user experience in text streaming services and to formally define and optimize QoE using a preemptive scheduling approach.
\name's scheduler can be integrated to most existing LLM serving systems.

\paragraph{Video Streaming and QoE.}
QoE in text streaming draws inspiration from QoE in video streaming~\cite{video-quality-engagement-sigcomm11,video-qoe-quest-hotnets12,video-control-plane-sigcomm12,pytheas-nsdi17,sensei-nsdi21} but encounters unique challenges and has a different QoE definition. 
While video streaming is primarily influenced by network conditions~\cite{qoe-modeling-http-access19}, text streaming is mainly constrained by GPU compute and memory~\cite{llm-survey-arxiv23}. 
Additionally, factors important to video streaming QoE include startup time, average bitrate, and buffering time ratio~\cite{cfa-nsdi16}.
While some have parallels in text streaming (\eg, startup time), others do not.
Our goal is to design a QoE metric tailored to text streaming and design a serving system that optimizes it.

\section{Conclusion}

In this work, we identify that user experience has thus far been overlooked in optimizing systems for LLM-based text streaming services.
This motivates us to define a QoE metric for text streaming services and build \name, a QoE-aware LLM serving system.
\name can deliver significantly higher QoE compared to existing systems, which also translates to being able to serve more concurrent requests while maintaining the same level of QoE, or reducing cost by saving GPU resources.
We hope that \name will encourage the community to dive deeper into understanding and optimizing user experience for text streaming services.

\label{EndOfPaper}

\bibliographystyle{plain}
\def\UrlBreaks{\do\/\do-}
\bibliography{andes-osdi25}

\clearpage

\appendix
\section{Alternative Scheduling Objectives}\label{sec:apdx-other-objectives}

In Section~\ref{sec:scheduling-problem-formulation}, we presented \name in terms of maximizing the average QoE across all requests.
However, text streaming services can have different quality goals under various deployment circumstances.
More importantly, our defined QoE metric and proposed solution can be seamlessly adapted to different QoE objectives.
In this section, we explore alternative scheduling objectives.

\parabf{Maximizing the Minimum QoE.}
To maximize the minimum QoE across all requests (i.e., max-min QoE), the gain (item value in knapsack) of request $i$ can be formulated as:
\begin{equation}
  \max(Q_{\min} - Q_{\textrm{wait}, i}, 0),
\end{equation}
where $Q_{\min}$ is the minimum QoE across all requests.
This function prioritizes requests that, if not served within \(\Delta t\), would further degrade the minimum QoE. 
By prioritizing these urgent requests, the overall QoE floor can be lifted, ensuring a more uniformly satisfying user experience.
 
\parabf{Maximizing the Number of Requests with Perfect QoE.}
To optimize the number of requests that achieve perfect QoE, the gain (item value in knapsack) of request $i$ can be formulated as:
\begin{equation}
  [\mathbb{1}(Q_{\textrm{serve, i}} = 1) - \mathbb{1}(Q_{\textrm{wait, i}} = 1) ] \cdot \mathbb{1}(Q_{\textrm{current}, i}=1),
\end{equation}
where $\mathbb{1}(\cdot)$ is 1 if the given condition is true and 0 otherwise, and $Q_{\textrm{current}, i}$ is the request's current QoE.
The intuition behind this approach is that
(1) there is no point in serving a request whose QoE is not perfect at the moment, and
(2) if a request with currently perfect QoE will degrade QoE if not served for $\Delta t$, the request must be prioritized.

\section{Modeling Token Generation Latency}\label{sec:apdx-modeling-token-generation-latency}

\begin{figure}
   \centering
   \includegraphics[trim=0 0 0 0,clip,width=.55\linewidth]{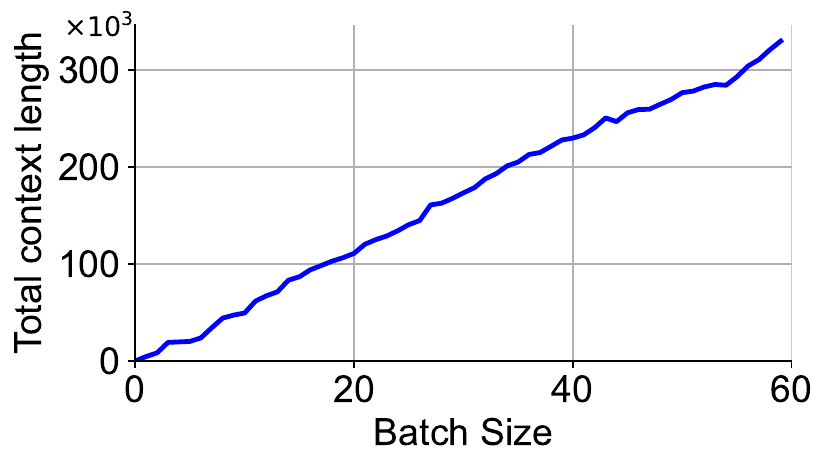} 
     \caption{Total context length distribution under different batch sizes using the Multi-Round ShareGPT dataset.}\label{fig:apdx-bs-context}
\end{figure}

In order to solve our knapsack formulation in Section~\ref{sec:scheduling-problem-formulation}, we need to be able to anticipate the QoE of a request after $\Delta t$ if served (i.e., $Q_{\textrm{serve}, i}$), which requires us to know the token generation latency, which is known to depend on batch size and the total number of tokens in the batch.

Figure~\ref{fig:apdx-bs-context} shows the relationship between the batch size and the total number of tokens across all requests in the batch (i.e., total context length).
It can be seen that batch size and total context length are nearly perfectly correlated, with Pearson correlation coefficient being 0.997.
Moreover, with the increase of batch size, the total context length is more predictable as it averages out the variance in individual request context lengths.
Therefore, we can drop total context length and estimate token generation latency simply as a function of batch size $B$.

\section{Dynamic Programming Solution}\label{sec:apdx-3d-dp}

\begin{algorithm}[t]
  \caption{Dynamic programming solution to Equation~\ref{eq:scheduling-optimization-problem}}\label{algo:scheduling-dp}
    \begin{algorithmic}[1]
    \Statex \textbf{Input:}\newline
    Number of requests $N$ and KV cache capacity $M$\newline
    Request context length array $l[N]$\newline
    Request QoE gain array $q[N]$\newline
    Target batch size $B$
    \Statex \textbf{Output:} Solution array $x[N]$.

    \State Initialize $dp[N+1][B+1][M+1]$ with $-\infty$
    \State Initialize $choice[N+1][B+1][M+1]$ with $0$
    \State $dp[0][0][0] = 0$

    \For{$i = 1$ to $N$}
        \For{$b = 0$ to $\min(i, B)$}
            \For{$m = 0$ to $M$}
                \If{$dp[i][b][m] < dp[i-1][b][m]$}
                    \Statex \hspace{2cm} \textcolor{blue}{$\triangleright$ Request $i$ is not served.}
                    \State $dp[i][b][m] = dp[i-1][b][m]$ 
                    \State $choice[i][b][m] = 0$
                \EndIf
 
                \If{$b \geq 1 \And m \geq l[i] $}
                    \If{$dp[i-1][b-1][m-l[i]] + q[i] > dp[i][b][m]$}
                        \Statex \hspace{2.5cm} \textcolor{blue}{$\triangleright$ Request $i$ is served.}
                        \State $dp[i][b][m] = dp[i-1][b-1][m-l[i]] + q[i]$
                        \State $choice[i][b][m] = 1$
                    \EndIf
                \EndIf
            \EndFor
        \EndFor
    \EndFor

    \State $Q_{\max} = \max(dp[N][B][:])$
    \State $m_{\textrm{current}} = \text{Index of } Q_{\max} \text{ in } dp[N][B]$
    \State $b_{\textrm{current}} = B$
    \State Initialize $x[N + 1]$ with zeros
    \For{$i = N$ \textbf{downto} $1$}
        \State $x[i] = choice[i][b_{\textrm{current}} ][m_{\textrm{current}}]$
        \If{$x[i] == 1$}
            \State $m_{\textrm{current}} = m_{\textrm{current}} - l[i]$
            \State $b_{\textrm{current}} = b_{\textrm{current}} - 1$
        \EndIf
    \EndFor

    \State \Return $x[1:]$ 

    \end{algorithmic}
\end{algorithm}

In Algorithm~\ref{algo:scheduling-dp}, we give a 3D dynamic programming solution to Equation~\ref{eq:scheduling-optimization-problem}.
The time complexity of the algorithm is $O(M \cdot N^2)$ as the largest batch size $B$ is $N$ in the worst case, and the problem needs to be solved for all feasible batch sizes $B$ to find the optimal set of requests to serve.
We note for clarity that our knapsack problem is \emph{weakly NP-Hard} and the 3D DP algorithm is \emph{not} polynomial time with respect to problem size (number of bits required to represent the problem).
That is, when the problem size (number of bits) is scaled in terms of the number of requests $N$ by adding more requests, runtime grows quadratically.
However, when the problem is scaled in terms of available memory $M$ by increasing the number of bits needed to represent $M$, the value of $M$ and thus algorithm runtime grows exponentially.
Therefore, the solution runs in \emph{pseudo-polynomial} time, which is effectively exponential time.
For more details on weak NP-Hardness and pseudo-polynomial runtime, we direct the reader to \cite{weak-np-completeness-wikipedia}.

\end{document}